\documentclass[iop]{emulateapj}

\usepackage{amssymb,amsmath,latexsym,graphics, graphicx,epsfig,multirow,comment,hyperref,appendix} 

\usepackage{natbib}
\usepackage{soul}
\usepackage{color}
\newcommand{\zinfall}{z_{\text{infall}}}
\newcommand{\Msun}{\text{ M}_{\odot}}
\newcommand{\msp}{\text{m}_{\text{sp}}}

\newcommand{\Metallicity}{\left[\text{Fe/H}\right]}
\newcommand{\Mh}{\text{M}_{\text{h}}}

\begin{document}
\nocite{*}

\title{Signatures of Kinematic Substructure in the Galactic Stellar Halo}
\author{Mariangela Lisanti}
\affiliation{Department of Physics, Princeton University, Princeton, NJ 08544}
\author{David N. Spergel}
\affiliation{Department of Astrophysical Science, Princeton University, Princeton, NJ 08544}
\author{Piero Madau}
\affiliation{Department of Astronomy \& Astrophysics, University of California Santa Cruz, Santa Cruz, CA 95064}
\affiliation{Center for Theoretical Astrophysics and Cosmology, Institute for Computational Science, University of Zurich, CH-9057 Zurich, Switzerland}

\date{\today}

\begin{abstract}
Tidal debris from infalling satellites can leave observable structure in the phase-space distribution of the Galactic halo.  Such substructure can be manifest in the spatial and/or velocity distributions of the stars in the halo.  
This paper focuses on a class of substructure that is purely kinematic in nature, with no accompanying spatial features.  To study its properties, we use a simulated stellar halo created by dynamically populating the Via Lactea II high-resolution $N$-body simulation with stars.  A significant fraction of the stars in the inner halo of Via Lactea share a common speed and metallicity, despite the fact that they are spatially diffuse.  We argue that this kinematic substructure is a generic feature of tidal debris from older mergers and may explain the detection of radial-velocity substructure in the inner halo made by the Sloan Extension for Galactic Understanding and Exploration.  The GAIA satellite, which will provide the proper motions of an unprecedented number of stars, should further characterize the kinematic substructure in the inner halo.  Our study of the Via Lactea simulation suggests that the stellar halo can be used to map the speed distribution of the local dark-matter halo, which has important consequences for dark-matter direct-detection experiments.\end{abstract}
\maketitle

\section{Introduction}

The process of galaxy formation alters the phase-space distribution of the dark matter (DM) and stellar components of the Milky Way (MW) halo.  The nature of the residual phase-space structure in the halo depends on the details of its formation history, and is sensitive to whether the galaxy formed from smooth collapse~\citep{Eggen:1962dj} or from the merger of many protogalactic fragments~\citep{Searle:1978gc}.  The $\Lambda$CDM paradigm currently provides the most well-motivated picture of MW formation, including both the dark and baryonic matter as basic ingredients.  In the $\Lambda$CDM framework, the MW halo forms from the hierarchical merging of smaller subhalos~\citep{White:1977jf}.   The subhalos are tidally disrupted as they fall into the host, and DM is torn off, along with stars that formed in the dense cores of the subhalos.  

Tidal remnants from a completely disrupted subhalo eventually come into equilibrium with the host halo.  An incomplete merging event, however, leaves tidal debris with phase-space structure distinguishable from that of the smooth equilibrated halo.  Dwarf galaxies are examples of infalling satellites that have not been completely disrupted.  These dwarfs  orbit about the MW, leaving tidal debris in their wake, especially near the turning points of their orbits where the tidal forces are strongest.  After tidal stripping, the debris exhibits distinctive structure in both position and velocity.  With time, the debris comes into equilibrium with the host, and any distinctive phase-space features are washed out.

The accretion events that build the MW stellar halo leave their imprint on the phase-space distribution of its constituent stars~\citep{Johnston:1996sb,Johnston:1997fv,Venn:2004hk}.  This structure persists for some time because stars exchange energy and momenta on timescales that are much longer than the age of the Galaxy~\citep{Helmi:2008eq}.  As a result, structure in the stellar halo serves as a fossil record of the MW's formation history and kinematic or spatial features may be indicative of one or more merger events.

The chemical composition of stars provides additional information about their origin~\citep{McWilliam:1997ua, Robertson:2005gv,Font:2005qs}.  The metal content is particularly indicative because iron is introduced into the interstellar medium from supernova explosions, and is thus related to the total integrated star formation.  The chemical properties of stars brought into the MW halo depend on the mass of their subhalo host because a more massive subhalo has had more time to form stars, and also retains more metal.  The stars that are deposited in the MW by such mergers are typically more metal-rich than those deposited earlier by smaller subhalos~\citep{Unavane:1995an}.  

Evidence for stellar substructure has been accumulating with the advent of large-scale surveys, such as the Sloan Digital Sky Survey (SDSS)~\citep{Fukugita:1996qt,Gunn:1998vh,York:2000gk,Smith:2002pca,Pier:2002iq,Ivezic:2004bf,Tucker:2006dv,Gunn:2006tw}, the Sloan Extension for Galactic Understanding and Exploration (SEGUE) \citep{AllendePrieto:2007ed,Lee:2007mf,Lee:2007ec,Yanny:2009kg}, the Spaghetti survey~\citep{Morrison:2000gp}, the Two Micron All Sky Survey~\citep{Skrutskie:2006wh}, the Quasar Equatorial Survey Team~\citep{1999AAS...195.1502C,2012MNRAS.427.3374M}, and the Radial Velocity Experiment~\citep{Steinmetz:2006qt, Zwitter:2008vc}.  The Sagittarius dwarf~\citep{Ibata:1994fv} is one of the most stunning examples of stellar substructure from an on-going accretion event; the dwarf is in the midst of a merger with the MW and the tidal stream it has left in its orbital wake has been mapped to amazing precision~\citep{Johnston:1995vd,Ivezic:2000ua, Yanny:2000ty, Ibata:2000pu,Vivas:2001dn,Majewski:2003ux}.  Many other examples of stellar substructure have been documented~\citep{Majewski:1996zz, Chiba:1997ta,Totten:1998,Helmi:1999uj, Helmi:1999ks,Totten:2000ab,Chiba:2000vu,Newberg:2001sx,Gilmore:2002jv,Helmi:2002iu, RochaPinto:2004ru, Grillmair:2006bd,Vivas:2006nh,Kepley:2007vx,Juric:2005zr, Klement:2008ws,Seabroke:2008,Watkins:2009im,Grillmair:2008fv,Starkenburg:2009nd,An:2009hj,Smith:2009kr, Klement:2009km, Schlaufman:2009jv,Harrigan:2010pd, Schlaufman:2011kf,Johnston:2012yh,Sheffield:2012yg}---\emph{e.g.}, the Monoceros ``Ring"~\citep{Yanny:2003zu, Ibata:2003di, Ivezic:2008wk}, the  Orphan Stream~\citep{Belokurov:2006ms, Grillmair:2006nx, Belokurov:2006kc}, the Virgo Stellar Stream~\citep{Duffau:2005ta,Zinn:2013saa}, and tidal tails near the \mbox{Pal 5}~\citep{Odenkirchen:2000zx,Rockosi:2002wu} and \mbox{NGC 5466}~\citep{Belokurov:2005ad,Grillmair:2006se} globular clusters.   

It is well-established that stellar substructure in position-space is washed out before substructure in velocity-space~\citep{Johnston:1997fv, Helmi:1999ks, Freeman:2002wq}.  A stream is an example of substructure that is coherent in both position- and velocity-space.  Compared to streams, purely kinematic substructure is older and thus provides a way to probe earlier mergers in the halo.   
In this paper, we focus on a particular class of purely kinematic substructure---termed ``debris flow"---whose constituents are not coherent in velocity, but still exhibit a characteristic \emph{speed}.  Debris flow is intermediate between an equilibrated halo and dynamically cold tidal stream.  

Debris flow has already been studied in the context of DM~\citep{Lisanti:2011as, Kuhlen:2012fz} and this work is the first to explore its signatures in the Galactic stellar halo.  In particular, we study a simulated stellar halo created by dynamically populating a DM-only $N$-body simulation with stars (Sec.~\ref{sec: VL2tagging}).  The simulated halo exhibits structure in velocity and metallicity that distinguishes it from the smooth-halo expectation (Sec.~\ref{sec: StellarFlows}).  In particular, the stars share a common speed and metallicity, determined by the mass and orbital properties of its host satellite, despite being spatially diffuse.

The observational signatures of debris flow, as well as its implications for current and future surveys, are described in Sec.~\ref{sec: ECHOS}.  A recent study of the SEGUE data~\citep{Schlaufman:2009jv,Schlaufman:2011kf} found evidence for  radial-velocity substructure in the inner halo.  The SEGUE study uses a very large sample of metal-poor stars and finds several high-confidence detections that are not associated with any known streams.  As we show, the SEGUE findings could potentially be explained by debris flow,  although confirmation will require more complete proper-motion measurements from the GAIA satellite~\citep{Perryman:2001sp, 2008IAUS..248..217L}.   

The mounting evidence for phase-space structure in the stellar halo supports the picture that the MW formed through hierarchical mergers.  In addition, it strongly suggests that the Galaxy's DM is not smoothly distributed in phase space.  Ideally, the identification of stellar substructure in the Solar neighborhood can be used to infer information about the local DM speed distribution, which is relevant for direct-detection experiments.  We conclude by discussing the implications of stellar debris flow for DM (Sec.~\ref{sec:DMImplications}).

\section{The Via Lactea II Stellar Halo}
\label{sec: VL2tagging}

Via Lactea-II (VL2) is a DM-only $N$-body simulation that employs about a billion $4.1\times10^3 \text{ M}_\odot$ particles to model the formation of a $M_{200}=1.9\times10^{12}$~M$_{\odot}$ MW-size halo and its substructure~\citep{Diemand:2008in, Zemp:2008gw}.  It is initialized at $z=104.3$ to a WMAP3 $\Lambda$CDM cosmology~\citep{Spergel:2006hy} and evolved to the present day.  Twenty-seven ``snapshots" of the simulation from \mbox{$z=27.54$} to \mbox{$z=0$}, spaced roughly 680 Myr apart, were analyzed in detail.  The subhalos in each snapshot were identified using the 6DFOF group finder~\citep{Diemand:2006ey}.  The progression of all 3200 subhalos with infall mass greater than $10^7$~M$_\odot$ was tracked from snapshot to snapshot to build the evolutionary history.  

To study the stellar counterpart of the VL2 DM halo, the simulation output was dynamically populated with stars~\citep{Rashkov:2011cq}.  The tagging prescription labeled the most tightly bound DM particles in each subhalo as stars, assigning a mass, $\msp$,\footnote{Just as one DM ``particle" in the simulation does not actually correspond to an elementary particle, one star ``particle" corresponds to a conglomerate of stars with total mass $\msp$.} and metallicity, $\Metallicity$, to each.  The tagging prescription was tuned to reproduce the luminosity function, the half-light radii, and the metallicity-luminosity relation of the MW's observed satellites.  We briefly review the tagging procedure here; for a more detailed discussion, see~\cite{Rashkov:2011cq} and references therein.  

The total stellar mass of the subhalo at infall, $\text{M}_*$, was assumed to follow the power-law
\begin{equation}
\frac{\text{M}_*}{\Mh} = 1.6\times10^{-5} \left( \frac{\Mh}{10^9 \Msun}\right)^{1.8} \, ,
\label{eq:totalmass}
\end{equation}  
where $M_h$ is the maximum mass of the subhalo's host.  $\text{M}_*/\Mh$ is the stellar formation efficiency and determines the satellite luminosity today.  The total stellar mass was distributed amongst the 1\% most-bound particles in the subhalo to get $\msp$.  This number determines the concentration of the stellar system at infall, and governs the amount of stellar material stripped at later times as well as the present-day structural properties of the surviving satellites. Tagging the 1\% highest total binding energy particles with stellar population provides a good fit to the distribution of half-light radii in MW dSphs~\citep{Rashkov:2011cq}.

The metallicity of each star was taken to be 
\begin{equation}
\Metallicity = -7.87 + 0.9\times \log \left( \frac{\Mh}{10^3 \Msun} \, \right).
\label{eq:RashkovMet}
\end{equation}
Stars in more massive subhalos have larger metallicities because their hosts retain more enriched material.  All stars in a subhalo are assigned the same metallicity.  Two processes---stellar mass loss from tidal stripping and the dimming of the stellar population with age---turn the above assumed scaling into the observed present-day luminosity-metallicity relation of~\cite{Kirby:2008ab}.  
%%%
\begin{figure}[b] %  figure placement: here, top, bottom, or page
	\begin{centering}
 	  \includegraphics[width=3.25in]{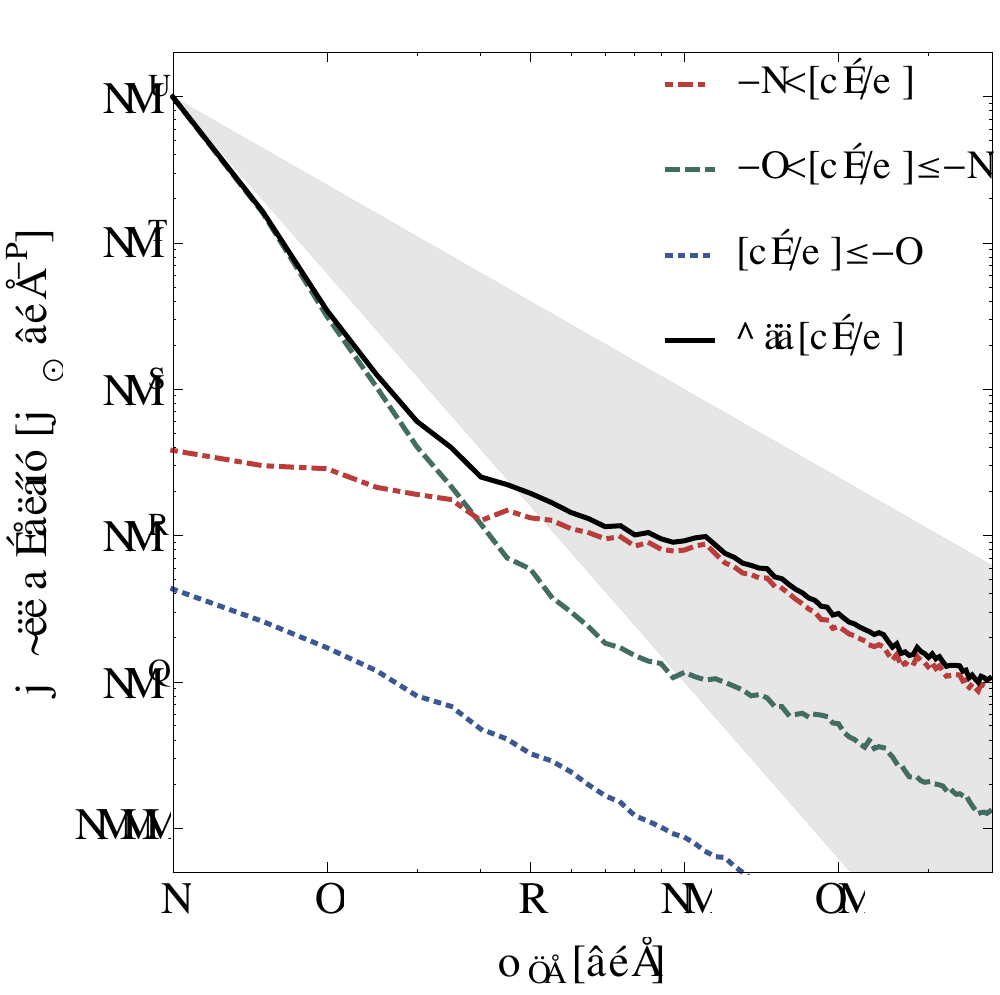} 
	  \end{centering}
	  \vspace{0.005cm}
   \caption{The mass density distribution of all unbound stars in VL2 as a function of galactocentric distance (solid black).  The individual contributions from stars with \mbox{$\Metallicity\leq$$-2$} (dotted blue), \mbox{$-2$$<\Metallicity\leq$$-1$} (dashed green), and  \mbox{$-1<\Metallicity$} (dot-dashed red) are also shown.  The shaded gray region indicates density fall-offs $\rho \propto r^{-\alpha}$ with $2<\alpha<4$, consistent with~\cite{Bell:2007ts}.}
   \vspace{-0.15cm}
   \label{fig:halodensity}
\end{figure}
%%%

%%
\begin{figure*}[tb] %  figure placement: here, top, bottom, or page
\begin{center}
   \includegraphics[width=6.5in]{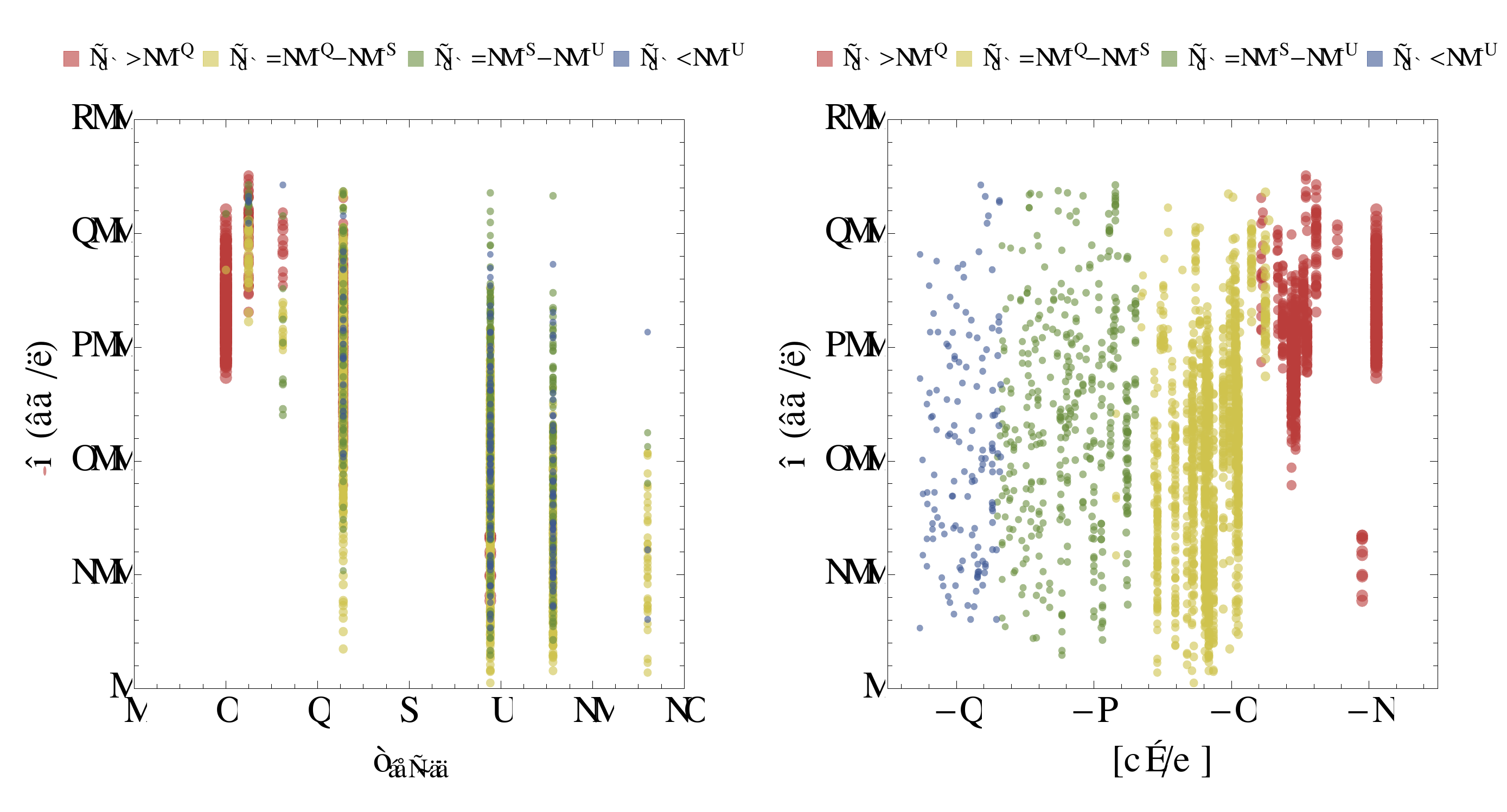}
   \end{center}
   \vspace{-0.2in}
   \caption{The galactocentric speeds of a random sampling of 3\% of the VL2 stars from \mbox{$R_\text{gc}=$5--15~kpc}, as a function of infall redshift $z_\text{infall}$ (left) and metallicity $\Metallicity$ (right).  Each point represents a star that constitutes a fraction $f_\text{GC}$ of the halo in this volume.  The stars are separated into four groups with $f_\text{GC} > 10^{-4}$ (red), 10$^{-4}$--10$^{-6}$ (yellow), 10$^{-6}$--10$^{-8}$ (green), $<10^{-8}$ (blue).  The discretization of infall redshifts is due to the snapshot sampling. }
   \vspace{0.2cm}
   \label{fig:scatter}
\end{figure*}

The stellar tagging procedure results in a list of 1.6~million stars in the VL2 host halo, along with the position, velocity, and metallicity of each.  About $31\%$ of the stars in the VL2 stellar halo are gravitationally bound to satellites at $z=0$ within \mbox{$R_\text{gc} < 100$ kpc}, where $R_\text{gc}$ is  galactocentric distance.  For \mbox{$R_\text{gc} =$ 5--15 kpc}, the bound fraction is only about 4$\%$.  The bound stars in the VL2 halo reproduce the sizes and brightness profiles of the observed dSph population, as well as their metallicities, velocity dispersions, and spatial distributions, as studied in~\cite{Rashkov:2011cq}.

The primary focus of this paper is the unbound population of stars in VL2.  The total mass and density distribution of the VL2 stellar halo reproduce the measured quantities from SDSS.  For example, as noted by~\cite{Rashkov:2011cq}, the $5\times10^8\Msun$ mass of the VL2 stellar halo compares well with observations in the same volume~\citep{Bell:2007ts}.  Figure~\ref{fig:halodensity} shows the mass density of all VL2 stars as a function of galactocentric distance (solid black).  The distribution exhibits a power-law behavior consistent with that observed by SDSS (indicated by the shaded gray band)~\citep{Bell:2007ts}.  Below \mbox{$R_\text{gc}\sim5\text{ kpc}$}, the VL2 density distribution is dominated by stars with \mbox{$-2<\Metallicity \leq -1$} (dashed green).  From \mbox{$R_\text{gc} = 5$--40 kpc}, the radial fall-off is determined by stars with \mbox{$-1 < \Metallicity$} (dot-dashed red).  Note that the maximum metallicity found in the VL2 stellar halo is \mbox{$\Metallicity=-0.95$}.  

Figure~\ref{fig:scatter} shows the galactocentric speeds of a random sampling of 3\% of the stars in VL2 from \mbox{$R_\text{gc}$=5--15~\text{ kpc}}, as a function of infall redshift (left panel) and metallicity (right panel).  Each point denotes a single star ``particle"; its color and relative size indicates the fraction of the total stellar mass in the volume from \mbox{$R_\text{gc}$=5--15~kpc} ($f_\text{GC}$) that it constitutes.  The VL2 stars can be separated into approximately two populations based on their metallicities.  The first population arises  from small-mass subhalos that fell into the MW at $z_\text{infall} \gtrsim 4.5$ and is more metal-poor than the population arising from more massive progenitors that merged  at \mbox{$z_\text{infall} \lesssim 3$}.\footnote{In the right panel of Fig.~\ref{fig:scatter}, there is a tail of high-metallicity stars that extends to low speeds.  This contribution to the high-metallicity sample is subdominant and was accreted around $z_\text{infall} \sim 8$.}  
%%%
\begin{table}
\begin{center}
\begin{tabular}{c @{\hskip 20pt} c@{\hskip 20pt} c }
$z_\text{infall}$	&	$f_\text{GC}$	& $f_\text{IH}$ \\	
\hline
\hline
2.00		&	$8.5\times10^{-1}$    &	$8.1\times10^{-1}$\\	
2.50		&	$1.8\times10^{-2}$	&	$2.4\times10^{-2}$\\
3.24		&	$6.4\times10^{-3}$	&	$3.0\times10^{-3}$	\\
4.56		&	$9.7\times10^{-2}$	&	$1.4\times10^{-1}$	\\
7.77		&	$2.9\times10^{-2}$	&	$1.4\times10^{-2}$	\\
9.14		&	$1.1\times10^{-3}$	&	$5.8\times10^{-4}$	\\
11.2		&	$6.2\times10^{-5}$	& 	$4.8\times10^{-5}$ 	\\
12.7 		& $3.8\times10^{-9}$		&	$7.2\times10^{-9}$	\\
\end{tabular}
\end{center}
\vspace{0.15cm}
\caption{The fraction of stellar mass that fell into the VL2 Milky Way host at redshift $z_\text{infall}$.  $f_\text{GC}$ is the fraction within \mbox{5--15 kpc} of the Galactic center; $f_\text{IH}$ is the fraction of stellar mass within the inner halo, defined in~\eqref{eq:innerhalo}.  Over $\sim$80$\%$ of the stellar mass was accreted at $z_\text{infall} = 2$ in either volume.} 
\label{tab:snapshots}
\vspace{-0.15cm}
\end{table}
%%%

The late-accreting and more metal-rich population in the VL2 stellar halo dominates its mass density.  Table~\ref{tab:snapshots} summarizes the fraction of stellar mass between \mbox{$R_\text{gc} =$ 5--15 kpc} ($f_\text{GC}$) that fell in at redshift $z_\text{infall}$.  For comparison, Table~\ref{tab:snapshots} also provides the fraction ($f_\text{IH}$) of stellar mass within the inner VL2 halo, defined, as in~\cite{Schlaufman:2009jv}, to be
\begin{equation}
\text{Inner Halo} = \left\{
     \begin{array}{l} 
     	\vspace{5pt}
       |z| >  4 \text{ kpc}\\ \vspace{5pt}
       R_\text{gc} > 10 \text{ kpc} \\ 
       d_\text{hel} < 17.5 \text{ kpc} \, ,
     \end{array}
          \right. 
          \label{eq:innerhalo}
\end{equation}
where $z$ is the vertical distance from the Galactic plane and $d_\text{hel}$ is the distance from the Sun.  We will come back to this volume in Sec.~\ref{sub:RVsubstructure}, when we compare the VL2 stellar halo with observations of the inner MW halo.  

Nearly 97\% of the stellar mass from \mbox{$R_\text{gc}$=5--15 kpc} was accreted after \mbox{redshift 4.56}, with $\sim$85\% accreted around \mbox{redshift 2}.   Indeed, the vast majority of the stellar halo in VL2 originates from a subhalo of stellar mass $\text{M}_* = 8.8\times10^8 \text{ M}_\odot$ that fell in at $z=2$~\citep{Rashkov:2011cq}.  Using (\ref{eq:totalmass}) and (\ref{eq:RashkovMet}), we estimate that this subhalo has a total mass of $\Mh = 5\times10^{10} \text{ M}_\odot$ and its stellar constituents have $\Metallicity = -0.94$.  This is in agreement with the large stellar mass density observed for $\Metallicity\sim-1$ in the right panel of Fig.~\ref{fig:scatter}.   

\begin{figure}[b] %  figure placement: here, top, bottom, or page
\begin{center}
   \includegraphics[width=3.25in]{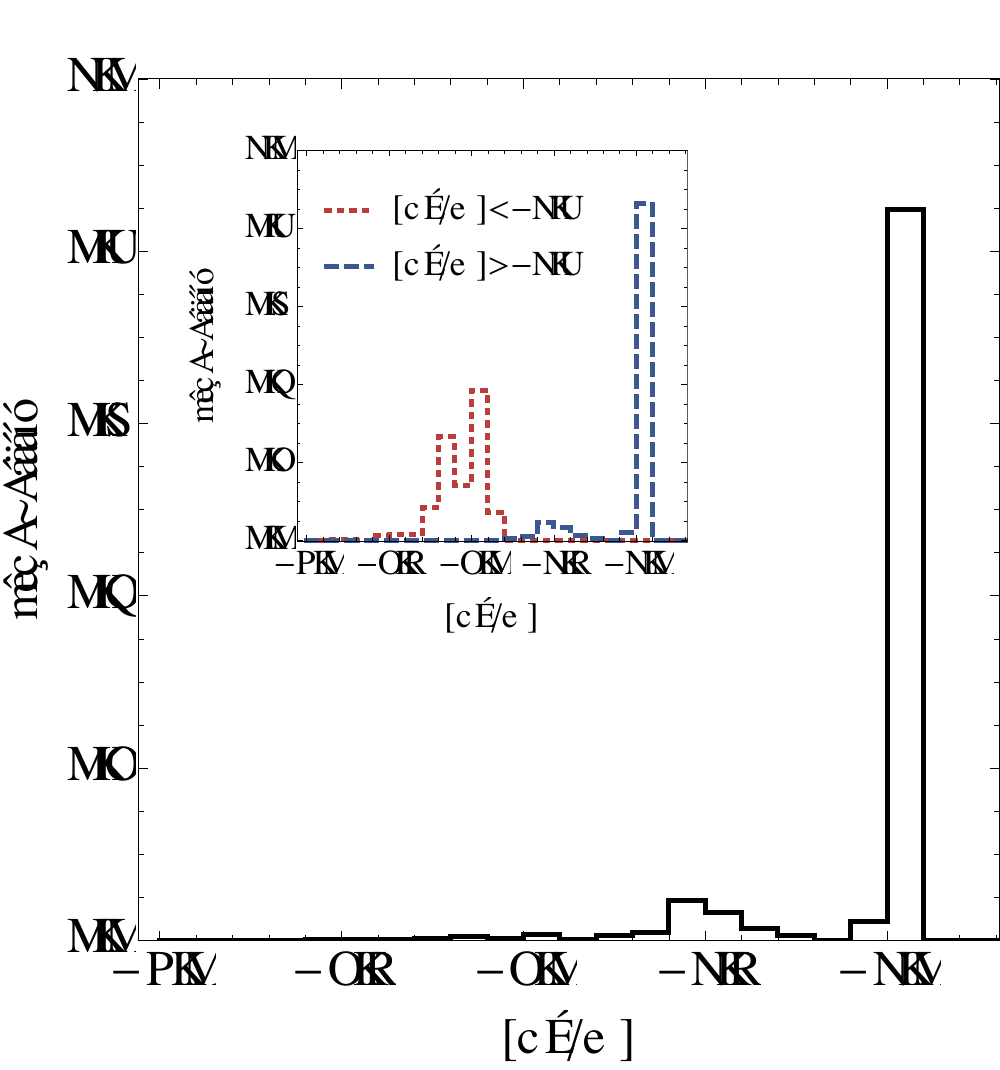}
   \end{center}
   \vspace{-0.1in}
   \caption{The metallicity distribution for VL2 stars with \mbox{$R_\text{gc}=5$--15 kpc}.  The inset shows the separately normalized distributions for the \mbox{$\Metallicity>-1.8$} (dashed blue) and \mbox{$\Metallicity<-1.8$} (dotted red) populations.} 
   \vspace{-0.15cm}
   \label{fig:velmetdist}
\end{figure}

From this point forward, we will divide the VL2 stars into two separate populations with 
\begin{equation}
\Metallicity < -1.8 \quad \text{and} \quad \Metallicity > -1.8  \, .
\end{equation}
A metallicity of $\Metallicity\sim-1.8$ is roughly where the average galactocentric speed of the stars changes from \mbox{$\sim$220 km/s} to \mbox{$\sim$330 km/s}.  The metallicity distribution for each population of VL2 stars  within \mbox{$R_\text{gc} =$ 5--15 kpc} is shown in Figure~\ref{fig:velmetdist}.  The more metal-rich population is narrowly peaked, with mean \mbox{$\Metallicity\sim-1.01$}.  The more metal-poor population has a broader distribution with mean \mbox{$\Metallicity\sim-2.07$}.  

There is a clear correlation between the kinematic and chemical properties of the stars in the VL2 halo.  Whether the actual stellar halo in the MW can be divided into two distinct populations---and what the relative fraction of each population is---depends entirely on its accretion history.  The stellar halo in VL2 is the manifestation of one possible accretion history and the observed features do not need to be generic.  Indeed, as was pointed out by~\cite{Schlaufman:2012ki}, the observed metallicity distributions in the Inner Halo are not consistent with VL2.  That being said, VL2 still serves as a valuable guide to better  understand the correlations between the kinematic and chemical properties of stars in the halo.

\section{Stellar Debris Flow}
\label{sec: StellarFlows}

\begin{figure*}[tb] %  figure placement: here, top, bottom, or page
\begin{center}
   \includegraphics[width=7in]{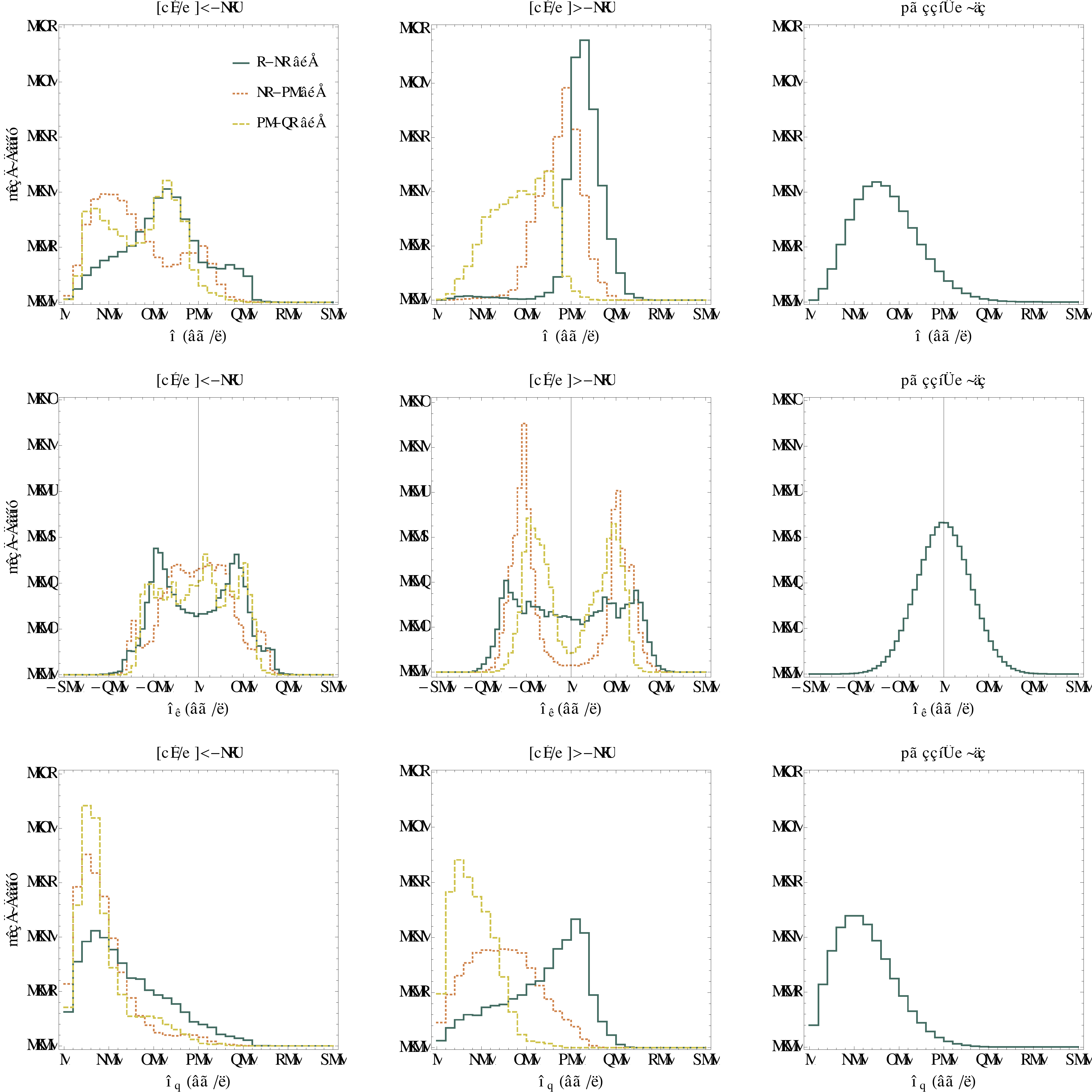}   
   \end{center}
      \vspace{-0.15in}
   \caption{Galactocentric speed (top row), radial velocity (middle row), and tangential velocity (bottom row) distributions for the VL2 stars with \mbox{$\Metallicity<-1.8$} (left column) and \mbox{$\Metallicity>-1.8$} (middle column).  The right column shows the corresponding distributions for the smooth inner halo, obtained from Monte Carlo using  (\ref{eq:rhoexpected}) and (\ref{eq:velexpected}).  The distributions are divided into radial bins with $R_\text{gc}=$ 5--15~kpc (solid green), 15--30~kpc (dashed orange), 30--45~kpc (dotted yellow).  There is a high-metallicity, high-speed contribution observed in VL2 that is not well-accounted for by the smooth-halo observations.}
   \vspace{0.2cm}
   \label{fig:velocitypanel}
\end{figure*}

Next, we analyze the metallicity and galactocentric velocity distributions of the stars in the VL2  halo.  The top row of Fig.~\ref{fig:velocitypanel} shows the galactocentric speed distributions for the VL2 halo stars with $\Metallicity<-1.8$ (left panel) and $\Metallicity>-1.8$ (middle panel).  In the \mbox{$R_\text{gc} =$ 5--15 kpc} radial bin, the $\Metallicity>-1.8$ population exhibits a narrow peak in galactocentric speed about the mean \mbox{$v\sim330$ km/s}.  The $\Metallicity<-1.8$ population has a much broader distribution in speed, but also exhibits a peak about \mbox{$v\sim230$ km/s}.  We focus primarily on the kinematic substructure observed in the more metal-rich population because the total mass density of these stars is much greater than that of the metal-poor population.

To compare the speed distributions obtained from VL2 to that expected for a smooth stellar halo, we simulate a mock catalog of halo stars whose distributions are consistent with observations of the inner MW halo (see~\cite{Ivezic:2013aja} for a review, and references therein).    The galactocentric position coordinates of the stars are assumed to follow a spherically-symmetric power-law with~\citep{Yanny:2000ty, Morrison:2000gp,Bell:2007ts}  
\begin{equation} 
\rho \propto r^{-3.5} \, .
\label{eq:rhoexpected}
\end{equation}
The galactocentric spherical velocity distribution is modeled as a multivariate normal distribution with mean and covariance matrix
\begin{equation}
\mu_{r,\theta,\phi} = \left( 
\begin{array}{c}
0 \\
0 \\
0 
\end{array} \right) \quad \quad
\Sigma_{r,\theta,\phi} = \left( 
\begin{array}{c c c}
120^2 	& 0 			& 0\\
0 			& 100^2	& 0\\
0 			& 0			& 100^2
\end{array} \right)  \, 
\label{eq:velexpected}
\end{equation}
(with velocity measured in units of km/s)~\citep{SommerLarsen:1996bn,Sirko:2003uq,Sirko:2003up,Xue:2008se}.  The mock catalog is constructed by selecting a star's galactocentric position and velocity coordinates from these distributions.  The density and velocity distributions in (\ref{eq:rhoexpected}) and (\ref{eq:velexpected}) match those used in~\cite{Schlaufman:2009jv}.%, a study that we will discuss in Sec.~\ref{sub:RVsubstructure}.  

The top right panel of Fig.~\ref{fig:velocitypanel} shows the galactocentric speed distribution for the mock star catalog.  Neither the $\Metallicity>-1.8$ nor the $\Metallicity<-1.8$ distributions resemble the smooth-halo expectation.  The more metal-rich population is significantly different, as it exhibits much higher speeds than what is expected for the smooth halo.  The metal-poor population bears more of a resemblance to the smooth-halo expectation, with lower speeds and a larger dispersion than its metal-rich VL2 counterpart, however it too exhibits a peaked feature indicative of kinematic substructure.  Note that the differences between the smooth halo and VL2 distributions cannot be accounted for by small variations in the parameters of (\ref{eq:rhoexpected}) and (\ref{eq:velexpected}).
%%%
%%%  

\begin{figure*}[htbp] %  figure placement: here, top, bottom, or page
   \centering
   \includegraphics[width=6.5in]{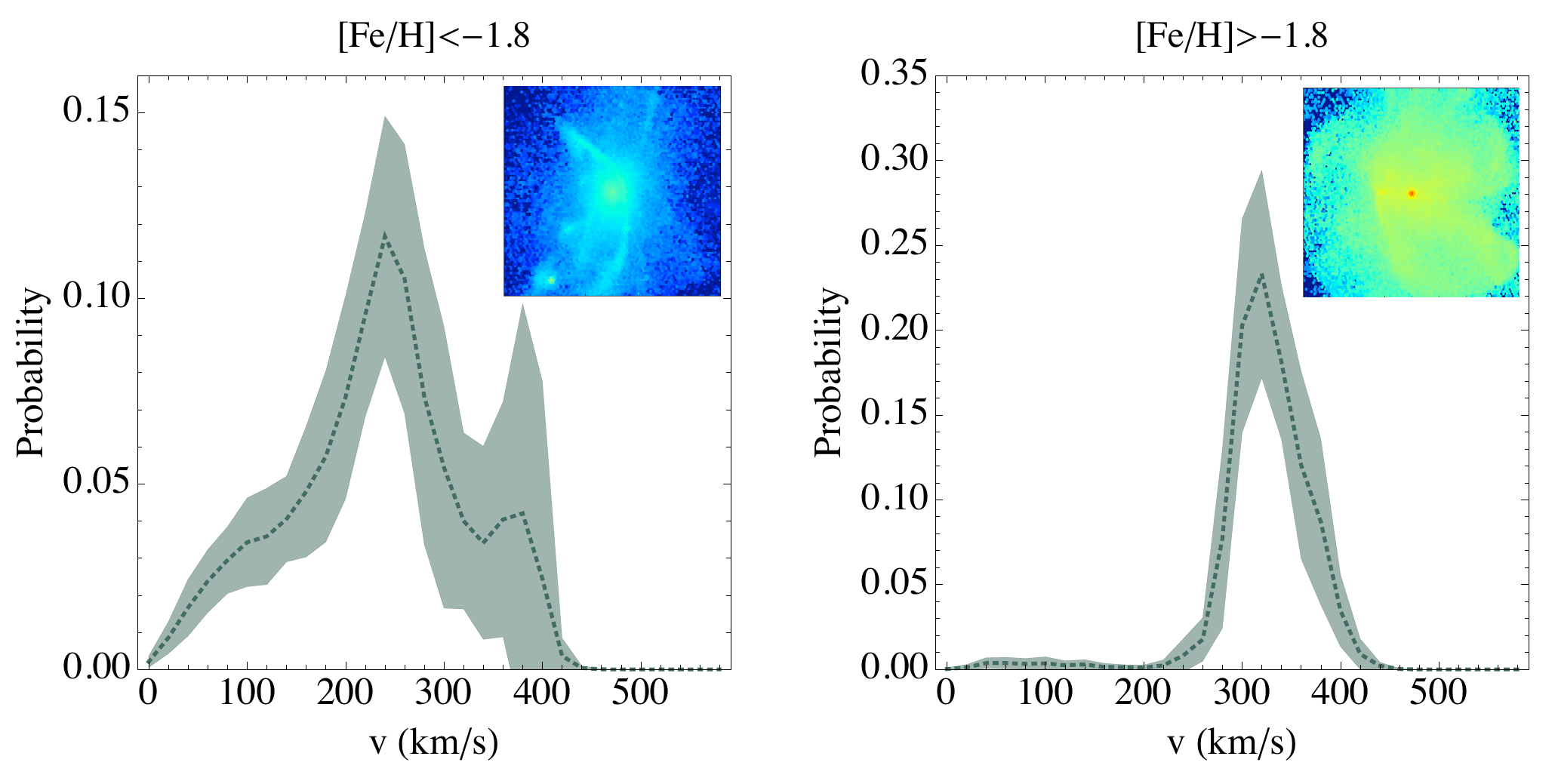} 
   \caption{The mean galactocentric speed distribution (dotted green) for 100 randomly selected spheres each with radius of 5 kpc, centered at \mbox{$R_\text{gc} = 10 \text{ kpc}$}.  The $\pm1\sigma$ region is indicated by the shaded green band.  The insets are projections of the respective mass density distributions in a \mbox{40 kpc$\times$40 kpc} box  positioned on the Galactic center.}
   \label{fig:vsampling}
\vspace{0.2cm}
\end{figure*}

The VL2 stellar halo has a high-metallicity, high-speed population of stars that is not consistent with observations of the smooth inner halo of the MW.  While the kinematic substructure is clear, there is no associated structure in position-space.  To illustrate this, we randomly select 100 spheres of radius \mbox{5~kpc} and centered at \mbox{$R_\text{gc}=10$ kpc}.  Figure~\ref{fig:vsampling} shows the mean galactocentric speed distributions (dotted line), as well as the $\pm1\sigma$ spread (shaded band) of these samples.  There is some variation in the speed distributions, however the peak at \mbox{$\sim$230 km/s} (for the $\Metallicity<-1.8$ population) and $\sim$330~km/s (for the $\Metallicity>-1.8$ population) are present over all the sampled regions.  This shows that this kinematic substructure is not associated with any spatial substructure, and is thus indicative of debris flow rather than a stream.

As a point of comparison, notice that the metal-poor population has another, smaller, peak at \mbox{$v\sim390 \text{ km/s}$}.  This peak is not present in all the sampled regions; at these speeds, a vanishing distribution is consistent to within one standard deviation.  As it turns out, the stars within this peak are associated with the stream that is visible in the inset of Fig.~\ref{fig:vsampling} (left panel).  When we look at the positions associated with the other kinematic structure in the VL2 stellar halo, there is no clear position-dependence as in this case. 

The speed distribution of stars with $\Metallicity>-1.8$ shifts to lower values as one moves to larger galactocentric distances, as illustrated in Fig.~\ref{fig:velocitypanel}.  Stars with \mbox{$R_\text{gc} =$ 15--30 kpc} have a mean speed of \mbox{280 km/s}, while those with \mbox{$R_\text{gc} =$ 30--45 kpc} have a mean speed of \mbox{190 km/s}.  The decrease in speed is set by energy conservation, as we now explain.

When a subhalo falls into the MW, it makes numerous orbits, losing mass from tidal stripping.  A rough estimate of the speed of the debris flow can be obtained from the energy conservation requirement:  
\begin{equation}
v^2\left(8.5\text{ kpc}\right) - v^2\left(D_\text{apo}^f\right) = 2 \left[\Phi\left(D_\text{apo}^f\right) - \Phi\left(8.5 \text{ kpc}\right) \right] \, ,
\label{eq:energyconservation}
\end{equation}
where $\Phi$ is the gravitational potential of the host halo and $D_\text{apo}^f$ is the distance of the final apocenter.  In~\cite{Kuhlen:2012fz}, it was found that the five most representative subhalos in the VL2 DM debris flow had a mean final apocenter distance of \mbox{$\langle D_\text{apo}^f \rangle = 59 \text{ kpc}$} and speed of \mbox{$\langle v_\text{apo} \rangle = 54 \text{ km/s}$}.  We will assume these values here for the stellar debris flow.
%%%
\begin{figure*}[t] %  figure placement: here, top, bottom, or page
\begin{center}
   \includegraphics[width=6in]{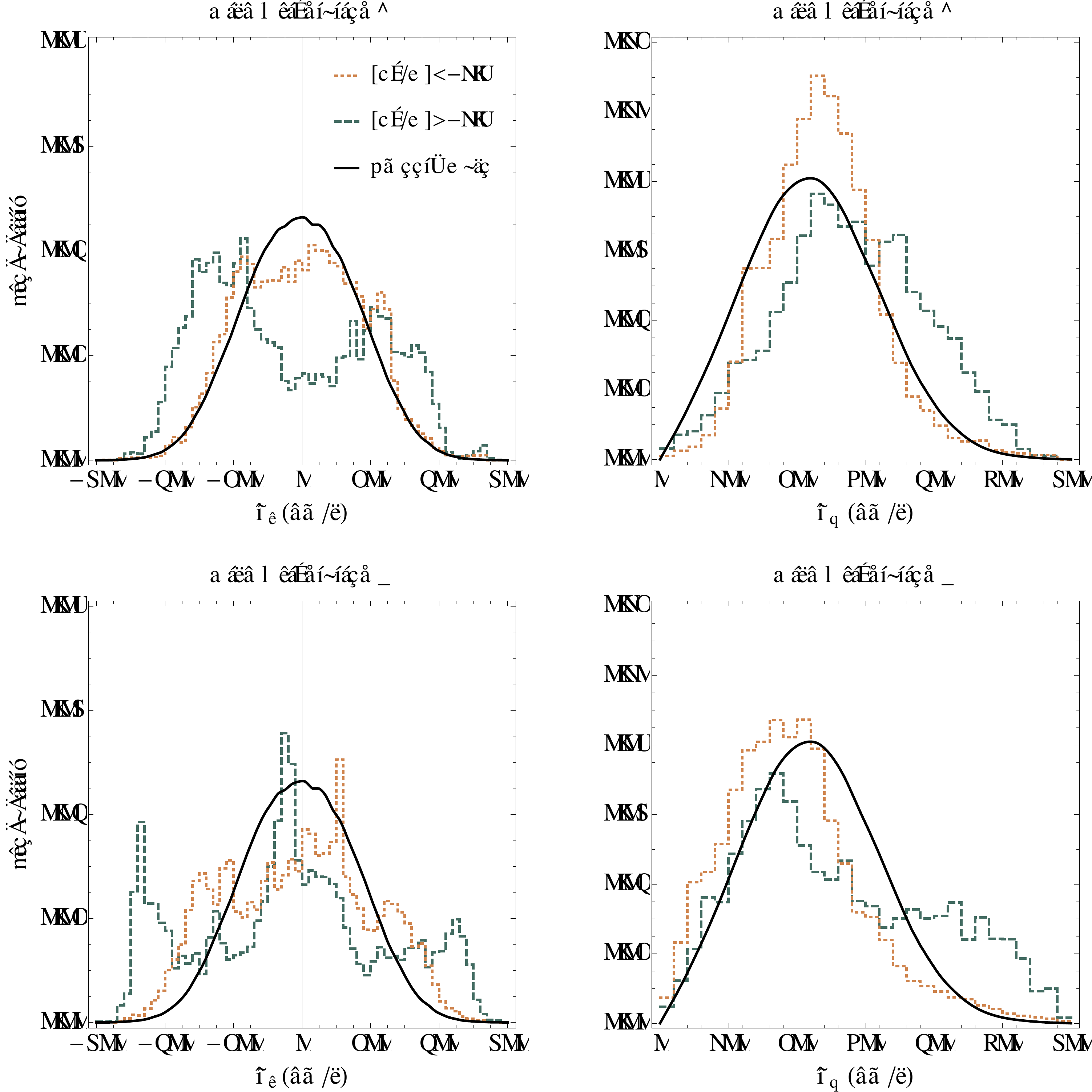} 
   \end{center}
   \caption{Radial (left column) and tangential (right column) velocities in the heliocentric frame for particles in the VL2 inner halo with $\Metallicity>-1.8$ (solid green) and $\Metallicity<-1.8$ (dotted orange).  The distributions are shown for two orientations of the Galactic disk in the VL2 host; because VL2 is a DM-only simulation, the choice of the plane orientation is arbitrary.  The smooth-halo expectation is shown by the solid black line.  Evidence for high-velocity behavior in the $\Metallicity>-1.8$ population is evident, either in the radial or tangential-velocity components, depending on the plane orientation. }
   \vspace{0.2cm}
   \label{fig:vlosvT}
\end{figure*}
Taking the NFW-like profile that best-fits the VL2 host halo~\citep{Diemand:2009bm}, we find that \mbox{$v(10 \text{ kpc})\sim360 \text{ km/s}$}, \mbox{$v(23 \text{ kpc})\sim280 \text{ km/s}$}, and \mbox{$v(38 \text{ kpc})\sim200 \text{ km/s}$}.  Therefore, if the tidal debris is dominated by a single satellite (or several satellites with similar properties at final apocenter), then its speed is primarily determined by the transfer of gravitational potential energy into kinetic energy as the satellite falls towards the Galactic center (GC).  This simple and intuitive picture reproduces the behavior of the high-metallicity population in Fig.~\ref{fig:velocitypanel} very well.
%%%
\begin{figure*}[t] %  figure placement: here, top, bottom, or page
      \includegraphics[width=7in]{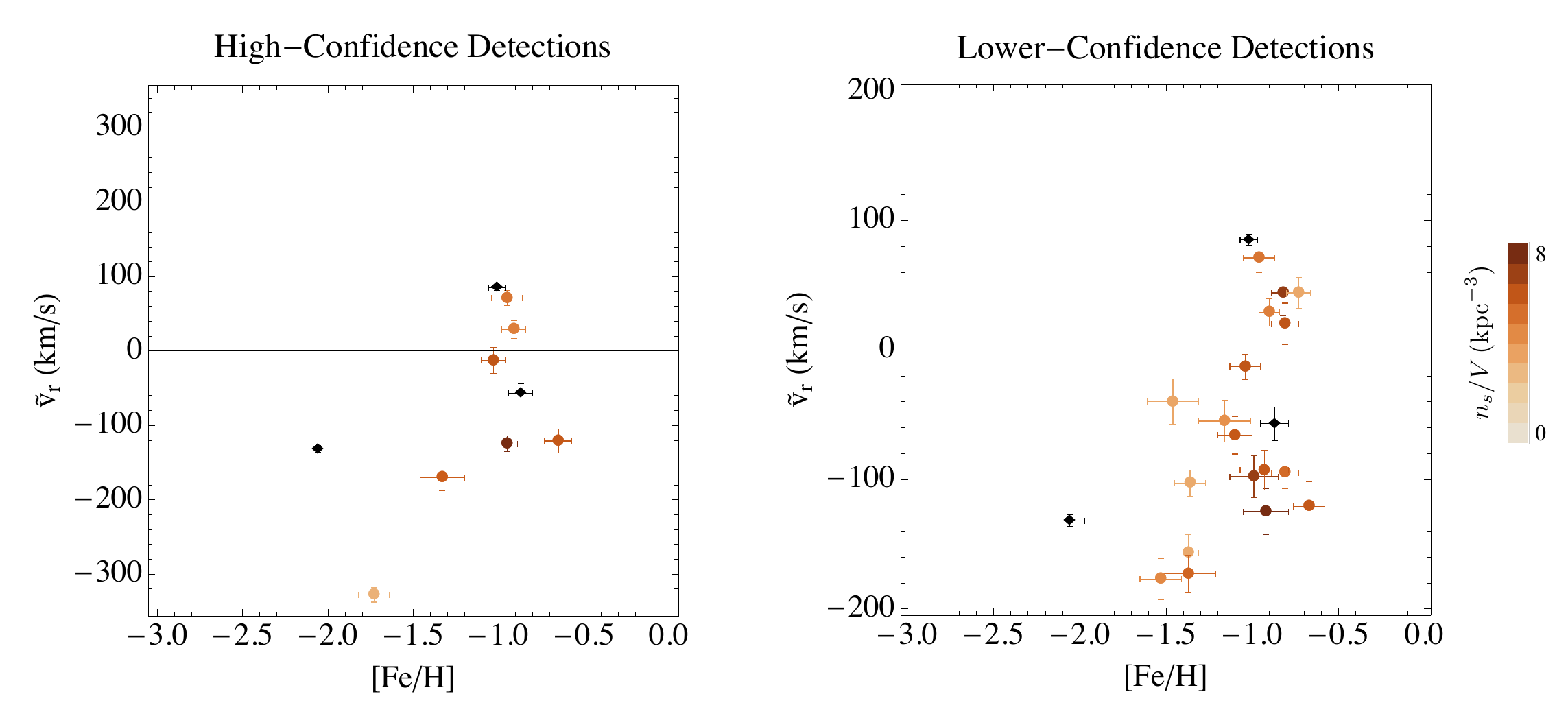} 
   \caption{Radial velocity in the heliocentric frame versus line-of-sight-averaged metallicity for the high-confidence (left) and lower-confidence substructure detections in~\cite{Schlaufman:2009jv,Schlaufman:2011kf}, indicated by the filled circles.  The color of each circle corresponds to the stellar number density of that substructure, $n_s/V$, where $n_s$ is the number of stars identified as being part of the substructure and $V$ is the volume scanned along that line-of-sight.  The black diamonds correspond to substructure detections that are associated with known streams.  Data taken directly from~\cite{Schlaufman:2011kf}.}
   \vspace{0.2cm}
   \label{fig:echos1}
\end{figure*}
%%%

The breakdown of the galactocentric velocity into its radial and tangential components depends on the orbital path of the infalling subhalo(s).  The second and third rows in Fig.~\ref{fig:velocitypanel} show the radial and tangential-velocity distributions, respectively, for the $\Metallicity<-1.8$ population (left panel), $\Metallicity>-1.8$ population (middle panel), and the smooth halo (right panel).  The kinematic substructure associated with the metal-rich population is primarily radial for \mbox{$R_\text{gc} \sim 15$--45 kpc} and becomes primarily tangential for \mbox{$R_\text{gc} \sim 5$--15 kpc}.  

Debris flow may be thought of as a collection of overlapping streams~\citep{Helmi:1999ks}.  In an infinitesimal phase-space volume, one will observe  $f_i(\textbf{x}, \textbf{v}) \, d^3\textbf{x} \, d^3\textbf{v}$ stars associated with stream $i$.  Because the stars in each stream are collisionless, their phase-space density is conserved as a function of time.  Therefore, as the stars in a given stream spread-out in position, they become more coherent in velocity.  However, the constituent stars share a common speed due to energy conservation.  Therefore, as each individual stream becomes more coherent in velocity, the dispersion in the average speed of its stars decreases.  The net result is that the dispersion in the stars' speeds (over all streams in the infinitesimal volume) also decreases.  The velocity direction of the stars varies from stream to stream, so the total debris (from all overlapping streams) is not coherent in total velocity, even though it is coherent in speed.

This simple picture leads to a concrete prediction for the observation of debris flow.  If the mass density of the debris from one or more subhalos of a given mass and infall redshift dominates the MW's inner halo, then one expects to observe stars that are spread out over large regions of sky (\emph{i.e.,} no spatial coherence), but share a common speed and metallicity.  Whether the kinematic substructure is evident in the stars' radial or tangential-velocity components depends on the properties of the satellites' orbits.  %If the debris is stripped as the subhalo(s) are falling towards/away from the GC, then the substructure is mainly in the radial-velocity component.  If it is stripped as the subhalo(s) are at a turning point, then the tangential-velocity component dominates.  

\section{Observational Evidence}
\label{sec: ECHOS}

Thus far, we have studied the properties of the VL2 debris flow in the Galactic frame.  To compare with observational studies, however, we need the distributions in the heliocentric frame.  To boost into this frame, we apply the following transformation:
\begin{equation}
\left(\tilde{v}_x, \tilde{v}_y, \tilde{v}_z\right) = (v_x - v_x^\odot, \, v_y - v_y^\odot, \, v_z - v_z^\odot) \, ,
\end{equation}
where $\left( v_x^\odot, v_y^\odot, v_z^\odot \right) = (-11.1, -232, 7.25) \text{ km/s}$~\citep{Schoenrich:2009bx}.  (We denote all heliocentric velocities with a tilde.)  It follows that the radial, $\tilde{v}_\text{r}$, and tangential, $\tilde{v}_\text{T}$, velocity components in the heliocentric frame are
\begin{eqnarray}
\tilde{v}_\text{r}&=& \tilde{v}_z \sin b - ( \tilde{v}_x \cos l + \tilde{v}_y \sin l) \cos b \\ \nonumber
\tilde{v}_\text{T} &=& \sqrt{ \tilde{v}_\text{b}^2 + \tilde{v}_\text{l}^2} \,,
\end{eqnarray}
where
\begin{eqnarray}
\tilde{v}_\text{b} &=& \tilde{v}_z \cos b + (\tilde{v}_x \cos l + \tilde{v}_y \sin l) \sin b\\ \nonumber
\tilde{v}_\text{l} &=& \tilde{v}_x \sin l - \tilde{v}_y \cos l \, ,
\end{eqnarray}
and $l$ and $b$ are the Galactic longitude and latitude, respectively.

Figure~\ref{fig:vlosvT} shows the radial and tangential velocities of the \mbox{$\Metallicity>-1.8$} (solid green) and \mbox{$\Metallicity<-1.8$} (dotted orange) populations in the VL2 inner halo, as defined by~(\ref{eq:innerhalo}).  Note that the orientation of the Galactic plane is arbitrary in VL2 because there is no baryonic disk, so the results for two disk orientations are presented. 

The velocity substructure persists in the heliocentric frame.  For Disk Orientation A, which is perpendicular to the z-axis, the radial velocity is peaked at \mbox{$\tilde{v}_r\sim\pm200$ km/s}.  This is quite different from the smooth-halo expectation (solid black line), which is distributed about zero.  In addition, the more metal-rich sample is skewed towards higher tangential velocities than its smooth-halo counterpart.  The kinematic substructure is primarily radial for Disk Orientation A.  This changes, however, with the orientation of the disk plane.  For example, when the disk is rotated by 90$^\circ$ (Disk Orientation B), the radial-velocity structure is still present, but suppressed, and the high-$\tilde{v}_\text{T}$ tail of the metal-rich distribution is more pronounced.  In this case, more stars have smaller radial velocities (explaining the peak at \mbox{$\tilde{v}_r\sim0$ km/s}), but larger tangential velocities.

The peaks in the radial-velocity distributions of Fig.~\ref{fig:vlosvT} are symmetric about zero, indicating that an equivalent amount of debris is being stripped from the satellites as they move towards and away from the GC.  One should keep in mind that this symmetry may be affected by the presence of the Galactic plane (not accounted for in VL2), which will affect the orbit of the satellites.  A complete characterization of this behavior requires studying debris flow in an $N$-body simulation that contains baryons.  

\subsection{Radial-Velocity Substructure in SEGUE}
\label{sub:RVsubstructure}

\cite{Schlaufman:2009jv} claim to detect radial-velocity substructure in the inner MW halo using a SEGUE catalog of 43,000 metal-poor main-sequence turnoff (MPMSTO) stars, 10,739 of which are located in the inner halo.  In general, MPMSTO stars are good tracers of the properties of the inner halo because they provide sufficient number density and luminosity.  SEGUE has 137 lines-of-sight, each designated by a unique latitude and longitude.  Each line-of-sight corresponds to a $\sim$$7^{\circ}$-square patch of sky and is separated from its nearest neighbor by $\sim$10$^{\circ}$--20$^{\circ}$.  As a result, one can assume that the lines-of-sight are independent of one another.  

\cite{Schlaufman:2009jv} obtained the radial-velocity distribution along each line-of-sight, and then compared it to the expected distribution for a smooth halo along that same line-of-sight.  The smooth halo was modeled using a mock star catalog that satisfied empirical measurements of the inner halo---the density and velocity distributions of the mock stars were sampled from distributions like (\ref{eq:rhoexpected}) and (\ref{eq:velexpected}).  Significant deviations in the measured line-of-sight velocities from those in the mock catalog point to the presence of kinematic substructure.  For a complete discussion of the statistical approaches used to identify deviations from the smooth-halo expectations, we refer the reader to~\cite{Schlaufman:2009jv}.  

\cite{Schlaufman:2009jv} report 10 high-confidence detections, in which no false-positives are expected.  Relaxing their statistical requirements, the authors also compile a list of 21 lower-confidence detections, of which less than three may be false positives.  The algorithm used to aggregate this collection of lower-confidence detections is more adept at picking out diffuse substructure along a line-of-sight, but at the expense of a greater false-positive rate.  Three each of the high-confidence and lower-confidence detections are associated with known streams~\citep{Schlaufman:2011kf}.  However, the remaining detections appear to be distributed roughly isotropically along all the lines-of-sight and do not exhibit a stream-like morphology.       

In a follow-up study,~\cite{Schlaufman:2011kf} found that these detections are chemically distinct from the smooth halo, strongly suggesting that they are the tidal remnants of a merger event.\footnote{The ECHOS are also less $\alpha$-enhanced than typical smooth-halo stars.  We do not focus on this here because the measured errors on [$\alpha$/Fe] are typically larger than those for $\Metallicity$.  In addition, the VL2 stellar tagging prescription does not include information on the [$\alpha$/Fe] of the stars.}  Figure~\ref{fig:echos1} shows the radial velocity of the high- and lower-confidence detections, versus their metallicity.  The detections with the largest number densities are clustered around $\Metallicity\sim-1$.  In general, the radial-velocity substructure detections in the SEGUE data are more metal-rich than typical halo stars, but more metal-poor than typical thick and thin-disk stars.  The kinematic and chemical properties of these stars are indicative of the tidal disruption of a merging satellite.  It is not clear whether the observed substructure is due to one or more merging satellites, however.  If it is only due to one merging event, then~\cite{Schlaufman:2011kf} estimate that the progenitor mass is at least \mbox{$10^9$ M$_\odot$} and that the merger took place around $z\lesssim0.5$, based on the metallicity and velocity dispersion of the detections.
  %%%
\begin{figure*}[t] %  figure placement: here, top, bottom, or page
      \includegraphics[width=7in]{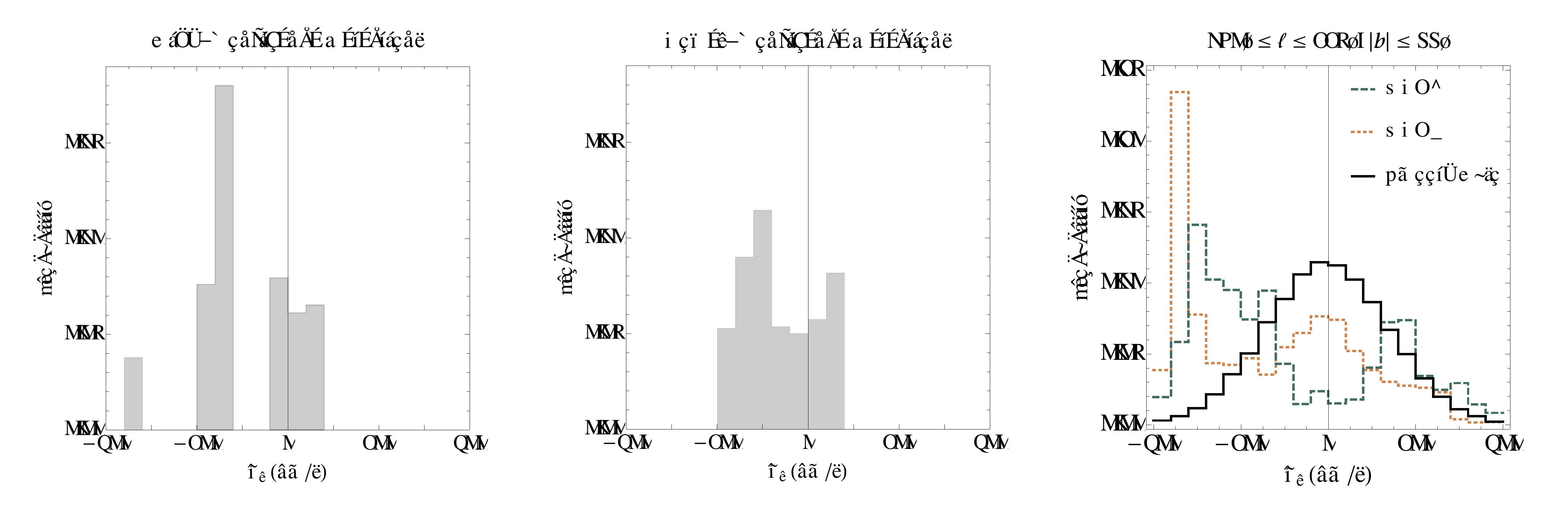} 
   \caption{The distribution of heliocentric radial velocities for the high- and lower-confidence substructure detections in~\cite{Schlaufman:2009jv,Schlaufman:2011kf} (left and middle panels, respectively).  The data is taken from~\cite{Schlaufman:2011kf}; the three known stream candidates are excluded.  Due to the selection function, most of these candidates are located between $130^\circ \leq l \leq 225^\circ$ and $|b| \leq 66^\circ$.  The right panel shows the predicted radial-velocity distribution for the smooth halo (solid black) in this region.  The corresponding distributions for Disk Orientation A and B for stars with $\Metallicity > -1$ in the VL2 inner halo are also shown (dashed green and dotted orange, respectively). }
   \vspace{0.2cm}
   \label{fig:echoshist}
\end{figure*}
%%%

To be consistent with debris flow, the substructure must be characterized by a coherent speed.  Only the heliocentric radial velocities of these detections was observed, so we start there to look for any patterns.  Figure~\ref{fig:echoshist} shows the distribution of the heliocentric radial velocities for the high- and lower-confidence detections (left and middle panels, respectively).  The expectation for the smooth halo is shown in the right panel (solid black), along with the distributions for stars with $\Metallicity>-1$ in the VL2 inner halo with Disk Orientation A (dashed green) and B (dotted orange).  Due to the selection function, the SEGUE detections are clustered around $130^\circ \leq l \leq 225^\circ$ and $|b| \leq 66^\circ$, so we add this additional requirement to the stars in the smooth halo and VL2 sample.  There are hints that the distribution of the lower-confidence detections is bimodal.  If this is the case, then the SEGUE detections would be more consistent with the VL2 distributions, than with the smooth-halo expectation.

\begin{figure}[b] %  figure placement: here, top, bottom, or page
      \includegraphics[width=3.25in]{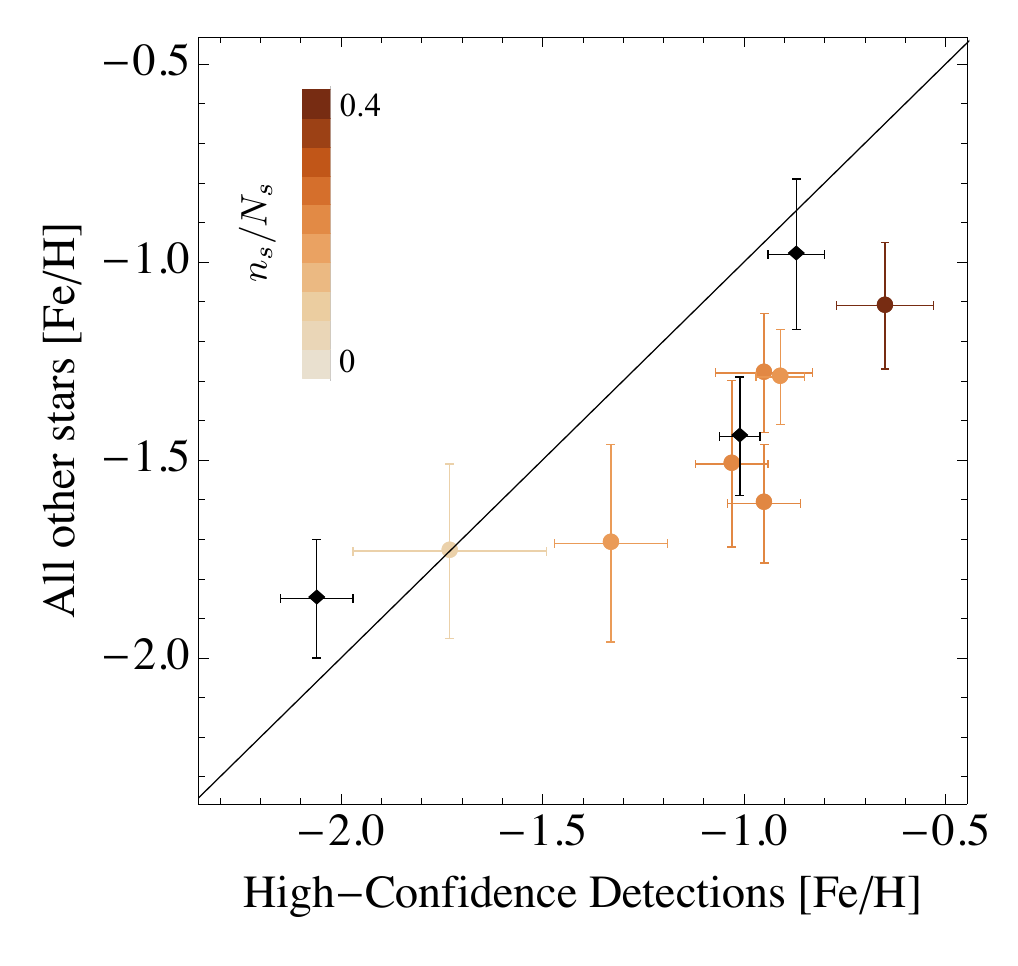} 
  \caption{The metallicity of the high-confidence detections in~\cite{Schlaufman:2009jv,Schlaufman:2011kf}, compared to the metallicity of the stars along the same line-of-sight that are not identified as being part of radial-velocity substructure (filled circles).  The color of the points indicates the $n_s/N_s$ along that line-of-sight, where $n_s$ is the number of stars in identified radial-velocity substructure and $N_s$ is the total number of other stars.  The black diamonds correspond to substructure detections that are associated with known streams.  The data is taken from~\cite{Schlaufman:2011kf}. }
   \vspace{0.2cm}
   \label{fig:echosmet1}
\end{figure}

As discussed earlier, the constituent stars of stellar debris flow share a common speed, set by the energetics of the satellite orbit.  The corresponding radial and tangential components of the stellar constituents depend on the orbital path of the satellite, but are always consistent with this speed.  The current data does suggest  that more substructure might indeed be present along the lines-of-sight.  Figure~\ref{fig:echosmet1} shows the metallicity of the stars labeled as being part of radial-velocity substructure and the metallicity of those that are not (along the same lines-of-sight), reproduced from~\cite{Schlaufman:2011kf}.  One cannot distinguish the latter set of stars from the smooth halo using the SEGUE data alone.  However, notice that the lines-of-sight that have a significant fraction of stars in radial-velocity substructure also have higher metallicities for all the other stars.  This correlation suggests that these other stars may not actually be part of the smooth halo, but are themselves in tangential-velocity substructure.  Our study of stellar debris flow in the VL2 simulated halo suggests that this should indeed be the case.  

%%%
%%%

While current observational data from SEGUE suggests the presence of local kinematic substructure in the stellar halo, a complete characterization requires the full three-dimensional velocities of the stars.  The GAIA satellite, which launched in December 2013, will provide high-precision radial and proper motions for an unprecedented number of stars in the MW~\citep{Perryman:2001sp, 2008IAUS..248..217L}.  This data will be critical for mapping the local stellar phase-space distribution.

If debris flow is present in the Solar neighborhood, one would expect to observe a group of stars that share a common metallicity and speed, but are not localized spatially.  The kinematic and chemical properties of the debris flow will depend on the subhalo(s) from which the stars were tidally stripped.  The fraction of the inner halo in kinematic substructure can be substantial---as evidenced by our study of VL2.  The GAIA satellite will provide the first opportunity to fully characterize this structure in detail, helping to decode information about older mergers in the MW halo and, potentially, the local DM distribution, as we now discuss.

\section{Dark Matter Implications \& Conclusions}
\label{sec:DMImplications} 

The presence of substructure in the phase-space distribution of the Galactic halo provides a fossil record of its evolutionary history.  In particular, the kinematic and chemical properties of the substructure hints at the time of the merger, as well as its size.  For instance, a recently accreted satellite may leave tidal debris in the form of streams, whose constituent stars exhibit structure in both position and velocity-space.  In contrast, the debris from an older merger event loses  coherence in position, even as it retains structure in velocity.  The debris from very old mergers is fully isotropized and indistinguishable in position and velocity from the host.  The metal content of the stellar debris is correlated with the mass of the merging satellite; more massive satellites lead to more metal-rich debris.

In this paper, we focus on older mergers whose tidal debris is only characterized by kinematic substructure---specifically, a coherent speed.  We refer to this class of substructure as ``debris flow" to distinguish it from tidal streams, which are coherent in both position and velocity.  We study the properties of stellar debris flow in the VL2 halo dynamically populated with stars.  The VL2 halo contains a subset of high-speed stars with metallicities $\Metallicity \sim -1$ that are distinct from the smooth halo.  Because the mass density distribution for this stellar subset exhibits no spatial features, it is an example of debris flow rather than a stream.

Having studied the stellar debris flow in VL2, we can now ask whether the DM in the simulated halo exhibits similar phase-space features.  A study of the DM debris flow in VL2 was completed in~\cite{Lisanti:2011as} and \cite{Kuhlen:2012fz}.  There, evidence was presented for kinematic substructure in the VL2 DM halo with the following properties.  First, it comprises $\sim$22\% of all the DM within \mbox{7.5--9.5 kpc} of the GC, but almost $\sim$80\% of the DM with speeds greater than $\sim$450 km/s.  Second, the density profile of the debris is consistent with that of the background halo within $\sim$50 kpc of the GC.  Third, the distribution of speeds is peaked $\sim$340 km/s, and the velocities are not coherent.  

The speed distribution for the DM debris flow in VL2 is remarkably similar to that of its stellar counterpart.  (Compare, for example, Fig.~\ref{fig:velocitypanel} of this work with Fig.~2 of~\cite{Lisanti:2011as}.)  In addition, both the DM and stellar debris appear to have similar origin.  The DM debris flow in VL2 is dominated by tidal debris from $\sim$5 satellites with infall time $\zinfall\sim2$ and mass $\sim10^9$--$10^{10}$ M$_\odot$.  The stellar halo in VL2 is dominated by a subhalo of similar mass that fell in at $z=2$.  This suggests that the stars are indeed a good tracer for the underlying DM distribution in the dynamically populated stellar halo of VL2.  

To see whether the features of the VL2 debris flow hold more generically requires repeating this analysis with other $N$-body simulations.  The speed and overall density of the debris flow depends on the merger history, which varies between simulations.  One would want to understand, for instance, the range of possibilities allowed for a set of realistic accretion histories.  In addition, a study of DM and stellar debris flow in a simulation that contains baryons is crucial for understanding how feedback mechanisms that inject energy redistribute the DM.  We expect that the tagging procedure used for VL2 should fail in regions dominated by baryons, such as near the GC or in the Galactic plane.  It has been shown, for example, that such tagging prescriptions can induce systematic uncertainties in the concentration and density structure of halos~\citep{Bailin:2014qfa}.  Regarding the properties of debris flow specifically, the presence of baryons can affect the orbits of subhalos as they pass through baryon-rich areas, which could change the properties of the flow.  To better understand the effect of baryons, we plan to study debris flow in Eris, a full hydrodynamic simulation that includes the effects of star formation and feedback~\citep{Guedes:2011ux,Kuhlen:2013tra}.  

Identifying the presence of stellar debris flow in the Solar neighborhood through its kinematic behavior and chemical abundance would have profound implications for DM searches.  The presence of local kinematic substructure would strongly suggest that a fraction of the local halo is comprised of debris from a merging satellite.  This would indicate that there are more high-speed DM particles in the solar neighborhood than would be expected for a fully equilibrated halo.  It can explain, for instance, deviations from the Maxwell-Boltzmann distribution along the high-velocity tail, as observed in numerical simulations~\citep{Fairbairn:2008gz,Vogelsberger:2008qb, Kuhlen:2009vh,Lisanti:2010qx}.  
These deviations can affect the annual modulation spectrum at direct-detection experiments~\citep{Freese:2012xd}, as discussed in detail in~\cite{Kuhlen:2012fz}.  In particular, debris flow may result in a larger modulation amplitude at high recoil energies (than would be expected for a Maxwell-Boltzmann distribution), especially for DM with large minimum scattering thresholds. 

In Sec.~\ref{sec: ECHOS}, we discussed evidence from SEGUE for radial-velocity substructure in the stellar halo.  The stars that are identified as being part of this kinematic substructure have distinctive metallicities ($\Metallicity\sim-1$) from the smooth halo and are isotropically distributed.  The morphology of the detections is consistent with debris flow.  A more complete understanding would require knowing the full proper motions of the stars, which will be provided by the upcoming GAIA satellite.  If the SEGUE results are confirmed, then they would provide evidence that a significant fraction of the local halo is not in equilibrium.  This would strongly suggest that, like the stellar debris in the halo, the local DM is in kinematic substructure as well.

\section*{Acknowledgments}
We thank M.~Kuhlen and V.~Rashkov for collaboration at the start of this work and for providing the stellar catalog.  We also thank J.~Bovy, S.~Gardner, A.~Kepley, K.~Johnston, H.~Morrison, C.~Rockosi, and M.~Valluri for helpful discussions.  This research was supported in part by the NSF under grants PHY11-25915 and  PHY-1066293.  Support for this work was also provided by the NSF through grants OIA-1124453 and AST-1229745 and by NASA through grant NNX12A587G(P.M.).\\

\vspace{0.3in}
\twocolumngrid
\def\bibsection{} 
\bibliography{starsBibFile}

\begin{thebibliography}{}
\expandafter\ifx\csname natexlab\endcsname\relax\def\natexlab#1{#1}\fi

\bibitem[{Allende~Prieto {et~al.}(2008)Allende~Prieto, Sivarani, Beers,
  {et~al.}}]{AllendePrieto:2007ed}
Allende~Prieto, C., Sivarani, T., Beers, T., {et~al.} 2008, Astron.J., 136,
  2070

\bibitem[{An {et~al.}(2009)An, Johnson, Beers, Pinsonneault, Terndrup,
  {et~al.}}]{An:2009hj}
An, D., Johnson, J.~A., Beers, T.~C., {et~al.} 2009, Astrophys.~J., 707, L64

\bibitem[{Bailin {et~al.}(2014)Bailin, Bell, Valluri, Stinson, Debattista,
  {et~al.}}]{Bailin:2014qfa}
Bailin, J., Bell, E.~F., Valluri, M., {et~al.} 2014, Astrophys.~J., 783, 95

\bibitem[{Bell {et~al.}(2008)Bell, Zucker, Belokurov, Sharma, Johnston,
  {et~al.}}]{Bell:2007ts}
Bell, E.~F., Zucker, D.~B., Belokurov, V., {et~al.} 2008, Astrophys.~J., 680,
  295

\bibitem[{Belokurov {et~al.}(2006{\natexlab{a}})Belokurov, Evans, Irwin,
  Hewett, \& Wilkinson}]{Belokurov:2005ad}
Belokurov, V., Evans, N., Irwin, M., Hewett, P.~C., \& Wilkinson, M.
  2006{\natexlab{a}}, Astrophys.~J., 637, L29

\bibitem[{Belokurov {et~al.}(2007)Belokurov, Evans, Irwin, Lynden-Bell, Yanny,
  {et~al.}}]{Belokurov:2006kc}
Belokurov, V., Evans, N., Irwin, M., {et~al.} 2007, Astrophys.~J., 658, 337

\bibitem[{Belokurov {et~al.}(2006{\natexlab{b}})Belokurov, Zucker, Evans,
  Gilmore, Vidrih, {et~al.}}]{Belokurov:2006ms}
Belokurov, V., Zucker, D., Evans, N., {et~al.} 2006{\natexlab{b}},
  Astrophys.~J., 642, L137

\bibitem[{Chiba \& Beers(2000)}]{Chiba:2000vu}
Chiba, M., \& Beers, T. 2000, Astron.~J., 119, 2843

\bibitem[{Chiba \& Yoshii(1998)}]{Chiba:1997ta}
Chiba, M., \& Yoshii, Y. 1998, Astron.~J., 115, 168

\bibitem[{{Coppi} {et~al.}(1999){Coppi}, {Snyder}, \& {QUEST
  Collaboration}}]{1999AAS...195.1502C}
{Coppi}, P., {Snyder}, J., \& {QUEST Collaboration}. 1999, in Bulletin of the
  American Astronomical Society, Vol.~31, American Astronomical Society Meeting
  Abstracts, 1395

\bibitem[{Diemand {et~al.}(2006)Diemand, Kuhlen, \& Madau}]{Diemand:2006ey}
Diemand, J., Kuhlen, M., \& Madau, P. 2006, Astrophys.~J., 649, 1

\bibitem[{Diemand {et~al.}(2008)Diemand, Kuhlen, Madau, Zemp, Moore,
  {et~al.}}]{Diemand:2008in}
Diemand, J., Kuhlen, M., Madau, P., {et~al.} 2008, Nature, 454, 735

\bibitem[{Diemand \& Moore(2011)}]{Diemand:2009bm}
Diemand, J., \& Moore, B. 2011, Adv.~Sci.~Lett., 4, 297

\bibitem[{Duffau {et~al.}(2006)Duffau, Zinn, Vivas, Carraro, Mendez,
  {et~al.}}]{Duffau:2005ta}
Duffau, S., Zinn, R., Vivas, A.~K., {et~al.} 2006, Astrophys.~J., 636, L97

\bibitem[{Eggen {et~al.}(1962)Eggen, Lynden-Bell, \& Sandage}]{Eggen:1962dj}
Eggen, O., Lynden-Bell, D., \& Sandage, A. 1962, Astrophys.~J., 136, 748

\bibitem[{Fairbairn \& Schwetz(2009)}]{Fairbairn:2008gz}
Fairbairn, M., \& Schwetz, T. 2009, JCAP, 0901, 037

\bibitem[{Font {et~al.}(2006)Font, Johnston, Bullock, \&
  Robertson}]{Font:2005qs}
Font, A.~S., Johnston, K.~V., Bullock, J.~S., \& Robertson, B. 2006,
  Astrophys.~J., 638, 585

\bibitem[{Freeman \& Bland-Hawthorn(2002)}]{Freeman:2002wq}
Freeman, K., \& Bland-Hawthorn, J. 2002, Ann.Rev.Astron.Astrophys., 40, 487

\bibitem[{Freese {et~al.}(2013)Freese, Lisanti, \& Savage}]{Freese:2012xd}
Freese, K., Lisanti, M., \& Savage, C. 2013, Rev.~Mod.~Phys., 85, 1561

\bibitem[{Fukugita {et~al.}(1996)Fukugita, Ichikawa, Gunn, Doi, Shimasaku,
  {et~al.}}]{Fukugita:1996qt}
Fukugita, M., Ichikawa, T., Gunn, J., {et~al.} 1996, Astron.J., 111, 1748

\bibitem[{Gilmore {et~al.}(2002)Gilmore, Wyse, \& Norris}]{Gilmore:2002jv}
Gilmore, G., Wyse, R.~F., \& Norris, J.~E. 2002, Astrophys.~J., 574, L39

\bibitem[{Grillmair(2006)}]{Grillmair:2006nx}
Grillmair, C.~J. 2006, Astrophys.~J., 645, L37

\bibitem[{Grillmair(2009)}]{Grillmair:2008fv}
---. 2009, Astrophys.~J., 693, 1118

\bibitem[{Grillmair \& Dionatos(2006)}]{Grillmair:2006bd}
Grillmair, C.~J., \& Dionatos, O. 2006, Astrophys.~J., 643, L17

\bibitem[{Grillmair \& Johnson(2006)}]{Grillmair:2006se}
Grillmair, C.~J., \& Johnson, R. 2006, Astrophys.~J., 639, L17

\bibitem[{Guedes {et~al.}(2011)Guedes, Callegari, Madau, \&
  Mayer}]{Guedes:2011ux}
Guedes, J., Callegari, S., Madau, P., \& Mayer, L. 2011, Astrophys.~J., 742, 76

\bibitem[{Gunn {et~al.}(1998)Gunn, Carr, Rockosi, {et~al.}}]{Gunn:1998vh}
Gunn, J.~E., Carr, M., Rockosi, C., {et~al.} 1998, Astron.J., 116, 3040

\bibitem[{Gunn {et~al.}(2006)Gunn, Siegmund, Mannery, {et~al.}}]{Gunn:2006tw}
Gunn, J.~E., Siegmund, W.~A., Mannery, E.~J., {et~al.} 2006, Astron.J., 131,
  2332

\bibitem[{Harrigan {et~al.}(2010)Harrigan, Newberg, Newberg,
  {et~al.}}]{Harrigan:2010pd}
Harrigan, M.~J., Newberg, H.~J., Newberg, L.~A., {et~al.} 2010,
  Mon.~Not.~Roy.~Astron.~Soc., 405, 1796

\bibitem[{Helmi(2008)}]{Helmi:2008eq}
Helmi, A. 2008, Astron.~Astrophys.~Rev., 15, 145

\bibitem[{Helmi \& White(1999)}]{Helmi:1999ks}
Helmi, A., \& White, S.~D. 1999, Mon.~Not.~Roy.~Astron.~Soc., 307, 495

\bibitem[{Helmi {et~al.}(1999)Helmi, White, de~Zeeuw, \& Zhao}]{Helmi:1999uj}
Helmi, A., White, S.~D., de~Zeeuw, P.~T., \& Zhao, H.-S. 1999, Nature, 402, 53

\bibitem[{Helmi {et~al.}(2003)Helmi, White, \& Springel}]{Helmi:2002iu}
Helmi, A., White, S.~D., \& Springel, V. 2003, Mon.~Not.~Roy.~Astron.~Soc.,
  339, 834

\bibitem[{Ibata {et~al.}(1994)Ibata, Gilmore, \& Irwin}]{Ibata:1994fv}
Ibata, R., Gilmore, G., \& Irwin, M. 1994, Nature, 370, 194

\bibitem[{Ibata {et~al.}(2001)Ibata, Lewis, Irwin, Totten, \&
  Quinn}]{Ibata:2000pu}
Ibata, R., Lewis, G.~F., Irwin, M., Totten, E., \& Quinn, T.~R. 2001,
  Astrophys.~J., 551, 294

\bibitem[{Ibata {et~al.}(2003)Ibata, Irwin, Lewis, Ferguson, \&
  Tanvir}]{Ibata:2003di}
Ibata, R.~A., Irwin, M., Lewis, G., Ferguson, A., \& Tanvir, N. 2003,
  Mon.~Not.~Roy.~Astron.~Soc., 340, L21

\bibitem[{Ivezic {et~al.}(2012)Ivezic, Beers, \& Juric}]{Ivezic:2013aja}
Ivezic, Z., Beers, T.~C., \& Juric, M. 2012, Annu.~Rev.~Astron.~Astrophys., 50,
  251

\bibitem[{Ivezic {et~al.}(2000)Ivezic, Goldston, Finlator,
  {et~al.}}]{Ivezic:2000ua}
Ivezic, Z., Goldston, J., Finlator, K., {et~al.} 2000, Astron.~J., 120, 963

\bibitem[{Ivezic {et~al.}(2004)Ivezic, Lupton, Schlegel,
  {et~al.}}]{Ivezic:2004bf}
Ivezic, Z., Lupton, R.~H., Schlegel, D., {et~al.} 2004, Astron.Nachr., 325, 583

\bibitem[{Ivezic {et~al.}(2008)Ivezic, Sesar, Juric, {et~al.}}]{Ivezic:2008wk}
Ivezic, Z., Sesar, B., Juric, M., {et~al.} 2008, Astrophys.~J., 684, 287

\bibitem[{Johnston(1998)}]{Johnston:1997fv}
Johnston, K.~V. 1998, Astrophys.~J., 495, 297

\bibitem[{Johnston {et~al.}(1996)Johnston, Hernquist, \&
  Bolte}]{Johnston:1996sb}
Johnston, K.~V., Hernquist, L., \& Bolte, M. 1996, Astrophys.~J., 465, 278

\bibitem[{Johnston {et~al.}(2012)Johnston, Sheffield, Majewski, \&
  Sharma}]{Johnston:2012yh}
Johnston, K.~V., Sheffield, A.~A., Majewski, S.~R., \& Sharma, S. 2012,
  Astrophys.~J., 760, 95

\bibitem[{Johnston {et~al.}(1995)Johnston, Spergel, \&
  Hernquist}]{Johnston:1995vd}
Johnston, K.~V., Spergel, D.~N., \& Hernquist, L. 1995, Astrophys.~J., 451, 598

\bibitem[{Juric {et~al.}(2008)Juric, Ivezic, Brooks, {et~al.}}]{Juric:2005zr}
Juric, M., Ivezic, Z., Brooks, A., {et~al.} 2008, Astrophys.~J., 673, 864

\bibitem[{Kepley {et~al.}(2007)Kepley, Morrison, Helmi, Kinman, Van~Duyne,
  {et~al.}}]{Kepley:2007vx}
Kepley, A., Morrison, H.~L., Helmi, A., {et~al.} 2007, Astron.~J., 134, 1579

\bibitem[{Kirby {et~al.}(2008)Kirby, Simon, Geha, Guhathakurta, \&
  Frebel}]{Kirby:2008ab}
Kirby, E.~N., Simon, J.~D., Geha, M., Guhathakurta, P., \& Frebel, A. 2008,
  Astrophys.J., 685, L43

\bibitem[{Klement {et~al.}(2008)Klement, Fuchs, \& Rix}]{Klement:2008ws}
Klement, R., Fuchs, B., \& Rix, H.-W. 2008, Astrophys.~J., 685, 261

\bibitem[{Klement {et~al.}(2009)Klement, Rix, Flynn, Fuchs, Beers,
  {et~al.}}]{Klement:2009km}
Klement, R., Rix, H.-W., Flynn, C., {et~al.} 2009, Astrophys.~J., 698, 865

\bibitem[{Kuhlen {et~al.}(2012)Kuhlen, Lisanti, \& Spergel}]{Kuhlen:2012fz}
Kuhlen, M., Lisanti, M., \& Spergel, D.~N. 2012, Phys.~Rev., D86, 063505

\bibitem[{Kuhlen {et~al.}(2014)Kuhlen, Pillepich, Guedes, \&
  Madau}]{Kuhlen:2013tra}
Kuhlen, M., Pillepich, A., Guedes, J., \& Madau, P. 2014, Astrophys.~J., 784,
  161

\bibitem[{Kuhlen {et~al.}(2010)Kuhlen, Weiner, Diemand, Madau, Moore,
  {et~al.}}]{Kuhlen:2009vh}
Kuhlen, M., Weiner, N., Diemand, J., {et~al.} 2010, JCAP, 1002, 030

\bibitem[{Lee {et~al.}(2008{\natexlab{a}})Lee, Beers, Sivarani,
  {et~al.}}]{Lee:2007mf}
Lee, Y.~S., Beers, T.~C., Sivarani, T., {et~al.} 2008{\natexlab{a}}, Astron.J.,
  136, 2022

\bibitem[{Lee {et~al.}(2008{\natexlab{b}})Lee, Beers, Sivarani,
  {et~al.}}]{Lee:2007ec}
---. 2008{\natexlab{b}}, Astron.J., 136, 2050

\bibitem[{{Lindegren} {et~al.}(2008){Lindegren}, {Babusiaux}, {Bailer-Jones},
  {Bastian}, {Brown}, {Cropper}, {H{\o}g}, {Jordi}, {Katz}, {van Leeuwen},
  {Luri}, {Mignard}, {de Bruijne}, \& {Prusti}}]{2008IAUS..248..217L}
{Lindegren}, L., {Babusiaux}, C., {Bailer-Jones}, C., {et~al.} 2008, in IAU
  Symposium, Vol. 248, IAU Symposium, ed. W.~J. {Jin}, I.~{Platais}, \&
  M.~A.~C. {Perryman}, 217--223

\bibitem[{Lisanti \& Spergel(2012)}]{Lisanti:2011as}
Lisanti, M., \& Spergel, D.~N. 2012, Phys.~Dark~Univ., 1, 155

\bibitem[{Lisanti {et~al.}(2011)Lisanti, Strigari, Wacker, \&
  Wechsler}]{Lisanti:2010qx}
Lisanti, M., Strigari, L.~E., Wacker, J.~G., \& Wechsler, R.~H. 2011,
  Phys.~Rev., D83, 023519

\bibitem[{Majewski {et~al.}(1996)Majewski, Munn, \& Hawley}]{Majewski:1996zz}
Majewski, S.~R., Munn, J.~A., \& Hawley, S.~L. 1996, Astrophys.~J., 459, L73

\bibitem[{Majewski {et~al.}(2003)Majewski, Skrutskie, Weinberg, \&
  Ostheimer}]{Majewski:2003ux}
Majewski, S.~R., Skrutskie, M., Weinberg, M.~D., \& Ostheimer, J.~C. 2003,
  Astrophys.~J., 599, 1082

\bibitem[{{Mateu} {et~al.}(2012){Mateu}, {Vivas}, {Downes}, {Brice{\~n}o},
  {Zinn}, \& {Cruz-Diaz}}]{2012MNRAS.427.3374M}
{Mateu}, C., {Vivas}, A.~K., {Downes}, J.~J., {et~al.} 2012, \mnras, 427, 3374

\bibitem[{McWilliam(1997)}]{McWilliam:1997ua}
McWilliam, A. 1997, Ann.~Rev.~Astron.~Astrophys., 35, 503

\bibitem[{Morrison {et~al.}(2000)Morrison, Mateo, Olszewski, Harding,
  Dohm-Palmer, {et~al.}}]{Morrison:2000gp}
Morrison, H.~L., Mateo, M., Olszewski, E.~W., {et~al.} 2000, Astron.~J., 119,
  2254

\bibitem[{Newberg {et~al.}(2002)Newberg, Yanny, Rockosi,
  {et~al.}}]{Newberg:2001sx}
Newberg, H.~J., Yanny, B., Rockosi, C., {et~al.} 2002, Astrophys.~J., 569, 245

\bibitem[{Odenkirchen {et~al.}(2001)Odenkirchen, Grebel, Rockosi,
  {et~al.}}]{Odenkirchen:2000zx}
Odenkirchen, M., Grebel, E.~K., Rockosi, C.~M., {et~al.} 2001, Astrophys.~J.,
  548, L165

\bibitem[{Perryman {et~al.}(2001)Perryman, de~Boer, Gilmore,
  {et~al.}}]{Perryman:2001sp}
Perryman, M., de~Boer, K.~S., Gilmore, G., {et~al.} 2001, Astron.~Astrophys.,
  369, 339

\bibitem[{Pier {et~al.}(2003)Pier, Munn, Hindsley, Hennessy, Kent,
  {et~al.}}]{Pier:2002iq}
Pier, J.~R., Munn, J.~A., Hindsley, R.~B., {et~al.} 2003, Astron.J., 125, 1559

\bibitem[{Rashkov {et~al.}(2012)Rashkov, Madau, Kuhlen, \&
  Diemand}]{Rashkov:2011cq}
Rashkov, V., Madau, P., Kuhlen, M., \& Diemand, J. 2012, Astrophys.~J., 745,
  142

\bibitem[{Robertson {et~al.}(2005)Robertson, Bullock, Font, Johnston, \&
  Hernquist}]{Robertson:2005gv}
Robertson, B., Bullock, J.~S., Font, A.~S., Johnston, K.~V., \& Hernquist, L.
  2005, Astrophys.~J., 632, 872

\bibitem[{Rocha-Pinto {et~al.}(2004)Rocha-Pinto, Majewski, Skrutskie, Crane, \&
  Patterson}]{RochaPinto:2004ru}
Rocha-Pinto, H.~J., Majewski, S.~R., Skrutskie, M., Crane, J.~D., \& Patterson,
  R.~J. 2004, Astrophys.~J., 615, 732

\bibitem[{Rockosi {et~al.}(2002)Rockosi, Odenkirchen, Grebel,
  {et~al.}}]{Rockosi:2002wu}
Rockosi, C.~M., Odenkirchen, M., Grebel, E.~K., {et~al.} 2002, Astron.~J., 124,
  349

\bibitem[{Schlaufman {et~al.}(2009)Schlaufman, Rockosi, Beers, Bizyaev,
  Brewington, {et~al.}}]{Schlaufman:2009jv}
Schlaufman, K.~C., Rockosi, C.~M., Beers, T.~C., {et~al.} 2009, Astrophys.~J.,
  703, 2177

\bibitem[{Schlaufman {et~al.}(2012)Schlaufman, Rockosi, Lee, Beers,
  Allende~Prieto, {et~al.}}]{Schlaufman:2012ki}
Schlaufman, K.~C., Rockosi, C.~M., Lee, Y.~S., {et~al.} 2012, Astrophys.J.,
  749, 77

\bibitem[{Schlaufman {et~al.}(2011)Schlaufman, Rockosi, Lee, Beers, \&
  Prieto}]{Schlaufman:2011kf}
Schlaufman, K.~C., Rockosi, C.~M., Lee, Y.~S., Beers, T.~C., \& Prieto, C.~A.
  2011, Astrophys.~J., 734, 49

\bibitem[{Schoenrich {et~al.}(2009)Schoenrich, Binney, \&
  Dehnen}]{Schoenrich:2009bx}
Schoenrich, R., Binney, J., \& Dehnen, W. 2009, arXiv:0912.3693

\bibitem[{Seabroke {et~al.}(2008)Seabroke, Gilmore, Siebert,
  {et~al.}}]{Seabroke:2008}
Seabroke, G.~M., Gilmore, G., Siebert, A., {et~al.} 2008,
  Mon.~Not.~Roy.~Astron.~Soc., 384, 11

\bibitem[{Searle \& Zinn(1978)}]{Searle:1978gc}
Searle, L., \& Zinn, R. 1978, Astrophys.~J., 225, 357

\bibitem[{Sheffield {et~al.}(2012)Sheffield, Majewski, Johnston, Cunha, Smith,
  {et~al.}}]{Sheffield:2012yg}
Sheffield, A.~A., Majewski, S.~R., Johnston, K.~V., {et~al.} 2012,
  Astrophys.~J., 761, 161

\bibitem[{Sirko {et~al.}(2004{\natexlab{a}})Sirko, Goodman, Knapp, Brinkmann,
  Ivezic, {et~al.}}]{Sirko:2003up}
Sirko, E., Goodman, J., Knapp, G.~R., {et~al.} 2004{\natexlab{a}}, Astron.~J.,
  127, 899

\bibitem[{Sirko {et~al.}(2004{\natexlab{b}})Sirko, Goodman, Knapp, Brinkmann,
  Ivezic, {et~al.}}]{Sirko:2003uq}
---. 2004{\natexlab{b}}, Astron.~J., 127, 914

\bibitem[{Skrutskie {et~al.}(2006)Skrutskie, Cutri, Stiening, Weinberg,
  Schneider, {et~al.}}]{Skrutskie:2006wh}
Skrutskie, M., Cutri, R., Stiening, R., {et~al.} 2006, Astron.J., 131, 1163

\bibitem[{Smith {et~al.}(2002)Smith, Tucker, Kent, {et~al.}}]{Smith:2002pca}
Smith, J.~A., Tucker, D.~L., Kent, S.~M., {et~al.} 2002, Astron.J., 123, 2121

\bibitem[{Smith {et~al.}(2009)Smith, Evans, Belokurov, Hewett, Bramich,
  {et~al.}}]{Smith:2009kr}
Smith, M., Evans, N., Belokurov, V., {et~al.} 2009,
  Mon.~Not.~Roy.~Astron.~Soc., 399, 1223

\bibitem[{Sommer-Larsen {et~al.}(1997)Sommer-Larsen, Beers, Flynn, Wilhelm, \&
  Christensen}]{SommerLarsen:1996bn}
Sommer-Larsen, J., Beers, T., Flynn, C., Wilhelm, R., \& Christensen, P. 1997,
  Astrophys.~J., 481, 775

\bibitem[{Spergel {et~al.}(2007)Spergel, Bean, Dore, {et~al.}}]{Spergel:2006hy}
Spergel, D., Bean, R., Dore, O., {et~al.} 2007, Astrophys.~J.~Suppl., 170, 377

\bibitem[{Starkenburg {et~al.}(2009)Starkenburg, Helmi, Morrison, Harding, van
  Woerden, {et~al.}}]{Starkenburg:2009nd}
Starkenburg, E., Helmi, A., Morrison, H.~L., {et~al.} 2009, Astrophys.~J., 698,
  567

\bibitem[{Steinmetz {et~al.}(2006)Steinmetz, Zwitter, Siebert, Watson, Freeman,
  {et~al.}}]{Steinmetz:2006qt}
Steinmetz, M., Zwitter, T., Siebert, A., {et~al.} 2006, Astron.J., 132, 1645

\bibitem[{Totten \& Irwin(1998)}]{Totten:1998}
Totten, E., \& Irwin, M. 1998, Mon.~Not.~Roy.~Astron.~Soc., 294, 1

\bibitem[{Totten {et~al.}(2000)Totten, Irwin, \& Whitelock}]{Totten:2000ab}
Totten, E., Irwin, M., \& Whitelock, P. 2000, Mon.~Not.~Roy.~Astron.~Soc., 314,
  630

\bibitem[{Tucker {et~al.}(2006)Tucker, Kent, Richmond,
  {et~al.}}]{Tucker:2006dv}
Tucker, D., Kent, S., Richmond, M.~W., {et~al.} 2006, Astron.Nachr., 327, 821

\bibitem[{Unavane {et~al.}(1996)Unavane, Wyse, \& Gilmore}]{Unavane:1995an}
Unavane, M., Wyse, R., \& Gilmore, G. 1996, Mon.~Not.~Roy.~Astron.~Soc., 278,
  727

\bibitem[{Venn {et~al.}(2004)Venn, Irwin, Shetrone, Tout, Hill,
  {et~al.}}]{Venn:2004hk}
Venn, K.~A., Irwin, M., Shetrone, M.~D., {et~al.} 2004, Astron.~J., 128, 1177

\bibitem[{Vivas {et~al.}(2001)Vivas, Zinn, Andrews, Bailyn, Baltay,
  {et~al.}}]{Vivas:2001dn}
Vivas, A., Zinn, R., Andrews, P., {et~al.} 2001, Astrophys.~J., 554, L33

\bibitem[{Vivas \& Zinn(2006)}]{Vivas:2006nh}
Vivas, A.~K., \& Zinn, R. 2006, Astron.~J., 132, 714

\bibitem[{Vogelsberger {et~al.}(2009)Vogelsberger, Helmi, Springel, White,
  Wang, {et~al.}}]{Vogelsberger:2008qb}
Vogelsberger, M., Helmi, A., Springel, V., {et~al.} 2009,
  Mon.~Not.~Roy.~Astron.~Soc., 395, 797

\bibitem[{Watkins {et~al.}(2009)Watkins, Evans, Belokurov, Smith, Hewett,
  {et~al.}}]{Watkins:2009im}
Watkins, L., Evans, N., Belokurov, V., {et~al.} 2009,
  Mon.~Not.~Roy.~Astron.~Soc., 398, 1757

\bibitem[{White \& Rees(1978)}]{White:1977jf}
White, S.~D., \& Rees, M. 1978, Mon.~Not.~Roy.~Astron.~Soc., 183, 341

\bibitem[{Xue {et~al.}(2008)Xue, Rix, Zhao, {et~al.}}]{Xue:2008se}
Xue, X.~X., Rix, H.~W., Zhao, G., {et~al.} 2008, Astrophys.~J., 684, 1143

\bibitem[{Yanny {et~al.}(2003)Yanny, Newberg, Grebel, {et~al.}}]{Yanny:2003zu}
Yanny, B., Newberg, H.~J., Grebel, E.~K., {et~al.} 2003, Astrophys.~J., 588,
  824

\bibitem[{Yanny {et~al.}(2000)Yanny, Newberg, Kent, {et~al.}}]{Yanny:2000ty}
Yanny, B., Newberg, H.~J., Kent, S., {et~al.} 2000, Astrophys.~J., 540, 825

\bibitem[{Yanny {et~al.}(2009)Yanny, Rockosi, Newberg, {et~al.}}]{Yanny:2009kg}
Yanny, B., Rockosi, C., Newberg, H.~J., {et~al.} 2009, Astron.J., 137, 4377

\bibitem[{York {et~al.}(2000)York, Adelman, Anderson, {et~al.}}]{York:2000gk}
York, D.~G., Adelman, J., Anderson, J.~E., {et~al.} 2000, Astron.J., 120, 1579

\bibitem[{Zemp {et~al.}(2009)Zemp, Diemand, Kuhlen, Madau, Moore,
  {et~al.}}]{Zemp:2008gw}
Zemp, M., Diemand, J., Kuhlen, M., {et~al.} 2009, Mon.~Not.~Roy.~Astron.~Soc.,
  394, 641

\bibitem[{Zinn {et~al.}(2013)Zinn, Horowitz, Vivas, Baltay, Ellman,
  {et~al.}}]{Zinn:2013saa}
Zinn, R., Horowitz, B., Vivas, A., {et~al.} 2013, arXiv:1312.1602

\bibitem[{Zwitter {et~al.}(2008)Zwitter, Siebert, Munari, Freeman, Siviero,
  {et~al.}}]{Zwitter:2008vc}
Zwitter, T., Siebert, A., Munari, U., {et~al.} 2008, Astron.J., 136, 421

\end{thebibliography}

\end{document}